\documentstyle[psfig,12pt]{article}

\def\beq{\begin{equation}}
\def\eeq{\end{equation}}
\begin{document}


\begin{titlepage}

\begin{flushright}
hep-th/0110058
\end{flushright}

\begin{centering}

\vspace*{3cm}

{\Large\bf On the Spectrum of QCD(1+1) with $SU(N_c)$ Currents}

\vspace*{1.5cm}

{\bf Uwe Trittmann}
\vspace*{0.5cm}

{\sl Department of Physics\\
Ohio State University\\
Columbus, OH 43210, USA}

\vspace*{1cm}


\vspace*{2cm}

{\large Abstract}

\vspace*{1cm}

\end{centering}

Extending previous work, we calculate the 
fermionic spectrum of two-dimensional QCD (QCD$_2$) 
in the formulation with $SU(N_c)$
currents. Together with the results in the bosonic sector this
allows to address the as yet unresolved task of finding the 
single-particle states of this theory as a function of 
the ratio of the numbers of flavors and colors, $\lambda=N_f/N_c$, 
anew. 
We construct the Hamiltonian matrix in DLCQ 
formulation as an algebraic function of the harmonic resolution 
$K$ and the continuous parameter $\lambda$ 
in the Veneziano limit.
We find 
that the fermion momentum is a function of $\lambda$ in the discrete
approach. 
A universality, existing only in two dimensions, dictates that  
the well-known 't Hooft and large $N_f$ spectra be reproduced
in the limits $\lambda\rightarrow 0$ and $\infty$, which we confirm. 
We identify their single-particle content which is surprisingly the 
same as in the bosonic sectors.
All multi-particle states are classified in terms of their constituents.
These findings allow for an identification of the lowest single-particles
of the adjoint theory. While we do not succeed in interpreting this 
spectrum completely, evidence is presented for the conjecture
that adjoint QCD$_2$ has a bosonic and an independent 
fermionic Regge trajectory of single-particle states.  
\vspace{0.5cm}


\end{titlepage}
\newpage


\section{Introduction}

Two-dimensional QCD 
will remain an interesting model for strong interaction 
physics until a first principles calculation of 
the low-lying spectrum of four-dimensional QCD is available.
The theory with one flavor of fundamental fermions coupled to  
non-abelian gauge fields was
solved by 't Hooft in his seminal paper \cite{tHooft74b} 
in the limit of a large
number of colors $N_c$. It is the prime example for the solution of 
a confining gauge theory and exhibits one
Regge trajectory of non-interacting mesons, while not possessing dynamical 
gluonic degrees of freedom. 
The theory can also be solved when the number of fundamental 
(flavored) fermions $N_f$ is large. 
This is the abelian limit of the theory, and it comprises
a single meson with mass $g^2 N_f/\pi$ \cite{AFST98}. 
So far it has, however, proven impossible to solve the theory
with fermions in the adjoint representation. This is
unfortunate, because 
adjoint fermions simulate the transverse gluons of 
realistic four-dimensional QCD.
The latter has, of course, $N_f=3$ fermions in the 
fundamental representation of $SU(N_c=3)$, 
but it has been established for several theories 
\cite{tHooft74b,finiteN,AntonuccioPinsky98} 
that the large $N_c$ limit is often a good approximation.
The difficulties with solving adjoint QCD$_2$ 
can be traced to the fact that parton pair 
production is not suppressed by factors $1/N_c$, contrary
to the 't Hooft model. One therefore
expects a rich spectrum of multiple Regge trajectories.
Adjoint QCD$_2$
has been discussed in the literature for almost a decade
\cite{DalleyKlebanov93b}--\cite{Engelhardt2001}. 
Many interesting facets of this theory have been revealed, {\em e.g.}~a 
confining/screening transition with a linearly decreasing string tension
at vanishing fermion mass 
\cite{Smilga,ArmoniFrishmanSonnenschein98}, an exponential rise of 
the density of states with the bound state mass which is reminiscent of
string theory \cite{BhanotDemeterfiKlebanov93}, and the fact that
the theory becomes supersymmetric at a special value of the fermion mass
\cite{Kutasov94}.
Still, frustratingly little about the (single-particle) solutions 
of this theory is known. 

 
Using the framework of discretized light-cone quantization(DLCQ) 
\cite{PauliBrodsky85a}, 
the numerical eigenvalue spectrum of adjoint QCD 
has been obtained by Dalley and Klebanov \cite{DalleyKlebanov93b}. 
The results have been 
improved in Refs.~\cite{BhanotDemeterfiKlebanov93,GHK97}.
The asymptotic spectrum of the theory has been calculated by Kutasov 
in the continuum \cite{Kutasov94}. 
There are mainly two reasons which prohibit the extraction of the 
single-particle solutions from these results.   
Firstly, especially the numerical results are obscured by the fact
that the standard formulation in terms of fermionic operators contains 
many multi-particle states. Secondly, 
in large $N_c$ calculations one is used to  
identifying single-particle states with single-trace states since the 
work of 't Hooft \cite{tHooft74b}. It was recently
established that this is not necessarily correct if one deals with 
fields in the adjoint representation \cite{GHK97,AntonuccioPinsky98,UT}. 
This might have consequences for results derived with this assumption
\cite{BhanotDemeterfiKlebanov93,KutasovSchwimmer95,ArmoniSonnenschein95,KoganZhitnitzky}.
It seems therefore that not so much the lack of results
but their interpretation is the main obstacle for solving the theory. 
To improve the situation in both respects 
we will adopt a strategy which is special to massless theories in two
dimensions. 
In the massless case, the light-cone Hamiltonian can be written as a 
pure current-current interaction,
and its Hilbert space splits up into sectors of different representations of 
the current algebra. The formulation of the theory in terms of $SU(N_c)$
currents which form a Kac-Moody algebra 
is therefore a preferred choice which we will use it in the present work.  
In this formulation, to be described in Sec.~\ref{SecQCD}, 
many of the multi-particle states 
will be absent, because only two of the current blocks give rise to 
single-particle states \cite{KutasovSchwimmer95}. The bosonic 
states lie in the so-called current block of the identity which was 
considered in Ref.~\cite{UT}.
The adjoint block gives rise to the fermionic
bound-states, to be calculated here, 
which we need for the interpretation of the full
spectrum, because the bosonic spectrum contains multi-particle states
with fermionic constituents \cite{GHK97}.
We will use the framework of DLCQ to realize the the dynamical 
operators on a finite-dimensional Fock basis. It turns
out that the momentum operator plays a special role in the fermionic
sector, and we will describe 
this and other peculiarities in Sec.~\ref{SecSpecial}. 
In Sec.~\ref{SecHamiltonian} 
we will construct the fermionic 
light-cone Hamiltonian in terms of the discrete 
momentum modes of the currents. The Hamiltonian is an algebraic function
of the cutoff in current number and, most importantly, of the
ratio $\lambda=N_f/N_c$. 

This explicit $\lambda$ dependence of the Hamiltonian and the eigenvalue 
spectrum
allows us to exploit a universality 
existing only in two dimensions, as part of our strategy 
to elucidate the spectrum of adjoint QCD$_2$.
The universality established in Ref.~\cite{KutasovSchwimmer95}
assures that the massive spectrum and interactions
of two-dimensional gauge fields coupled to massless matter 
are largely independent 
of the representation of the matter fields, given they have the same 
chiral anomaly. 
All information on the matter representation 
beyond its Kac-Moody level is encoded in the massless sector of the theory.
There is, however, no strict factorization between massive and massless 
sectors, although 
the Hamiltonian is determined by the states from the massive sector only. 
In particular, massive states will have well-defined 
discrete symmetry quantum numbers only when accompanied by massless states 
\cite{KutasovSchwimmer95}.
It is clear that the universality can hold in two dimensions only.
In four dimensions massive and massless modes are known to be strongly 
interacting.
So far, this universality has been understood
in light-front quantization only.
The universality specifically predicts that the massive
spectrum of the Yang-Mills theory coupled to one adjoint $SU(N_c)$ fermion is
the same as the spectrum of the theory coupled to  
$N_f=N_c$ flavors of fundamental fermions. 
If this is true, we should 
obtain the 't Hooft spectrum in the limit of vanishing 
$\lambda$ and a single meson in the large $N_f$ limit in our numerical
calculations.
This exercise is performed in sections \ref{SecLargeNf} and \ref{SecTHooft} 
with the predicted result.
This confirms that the universality can be applied to the present case 
and provides a strong test of the numerics. 
In both limits the multi-particle states decouple 
and we succeed in describing the spectra in terms of their single-particle
content, thus classifying all multi-particle states by their constituents.
This helps us to understand the adjoint spectrum in Sec.~\ref{SecAdjoint}. 
While we are able to identify the low-lying single particle states and 
to construct some of the multi-particle states, 
a complete solution of the theory remains elusive. Motivated by 
empirical findings and an analysis of the spectrum at intermediate 
values of $\lambda$ in Sec.~\ref{SecInter}, we are led to the {\em conjecture}
that there are two Regge trajectories in the adjoint theory: a bosonic
and a fermionic one. We discuss the speculative character of these results, 
tests and possible improvements in the concluding 
Sec.~\ref{SecSummary}. 
In the appendix we display a calculation of the first corrections 
in $\lambda$ to the mass 
to the 't Hooft mesons. The agreement with the results of a recent 
perturbative analysis \cite{Engelhardt} provides further support for 
the usefulness of the present formulation of massless QCD$_2$ in terms of 
current operators.


\section{QCD in two dimensions}
\label{SecQCD}

The aim of the present work is to compute the massive spectrum
of $SU(N_c)$ Yang-Mills gauge fields coupled to massless fermions in 
some representation $r$ in two dimensions. 
The Veneziano limit, where both $N_f$ and $N_c$ are large,
is understood throughout.
A universality, existing only for massless {two-dimensional} gauge theories 
\cite{KutasovSchwimmer95}, predicts that the massive spectrum of the theory
is the same, whether one adjoint $SU(N_c)$ Majorana fermion 
or $N_f{=}N_c$ flavors of fundamental Dirac fermions are coupled to 
the gauge fields.
This means that we can formulate the theory in terms of adjoint fields, 
while interpreting the results in terms of fundamentals.
This gives us a continuous parameter at hand,
namely $\lambda{=}N_f/N_c$, which allows us to couple the Yang-Mills fields
to matter in different representations by simply altering its value,
while still keeping $N_f$ and $N_c$ large. 
This in turn gives deeper insight into the theory,
since the spectra in the limits $\lambda{\rightarrow}0$ ('t Hooft model)
and $\lambda{\rightarrow}\infty$ (large $N_f$ model)
are well 
understood,
whereas the single-particle content of the adjoint theory remains largely 
unknown.
Consequently, the main focus is on the case $\lambda{=}1$, 
while we will try to infer as much information
as possible from the 't Hooft and large $N_f$ models
by analyzing them in the formulation with current operators. 

In order to do so we have to derive the momentum and energy operators
in terms of currents rather than with fermionic operators. 
We consider the 
adjoint theory, but shall distinguish $N_f$ and $N_c$ 
throughout the derivation. As we saw, one can 
formally interpret the results at
different $N_f$ and $N_c$ as distinct theories. 
The Lagrangian
in light-cone coordinates $x^\pm=(x^0\pm x^1)/\sqrt{2}$, where $x^+$ plays the
role of a time, reads
\beq
{\cal L}=Tr[-\frac{1}{4g^2}F_{\mu\nu}F^{\mu\nu}+
i\bar{\Psi}\gamma_{\mu}D^{\mu}\Psi]
\eeq
where $\Psi=2^{-1/4}({\psi \atop \chi})$, 
with $\psi$ and $\chi$ 
being $N_c\times N_f$ matrices. The field strength is
$F_{\mu\nu}=\partial_{\mu}A_{\nu}-\partial_{\nu}A_{\mu}+i[A_{\mu},A_{\nu}]$,
and the covariant derivative is defined as $D_{\mu}=\partial_{\mu}
+i[A_{\mu},\cdot]$. 
We work in the light-cone gauge, $A^+=0$, which is consistent if we omit
the fermionic zero modes. The massive spectrum is not affected by this
omission \cite{Dave}. We use the convenient Dirac basis
$\gamma^0=\sigma_1$, $\gamma^1=-i\sigma_2$. 
The Lagrangian then becomes
\beq
{\cal L}=Tr\left[\frac{1}{2g^2}(\partial_-A^-)^2+i\psi^{\dagger}\partial_
+\psi+i\chi^{\dagger}\partial_-\chi-A^-J\right],
\eeq
with the current 
\begin{equation}
J^a_b(x^-)=2:\psi^{\dagger a}_c(x^-)\psi^c_b(x^-):.
\label{current}
\end{equation}
The use of both upper and lower indices is adopted as a reminder that
in general the indices are from different index sets. 
We can integrate out the 
non-dynamical component $A^-$ of the gauge field and obtain 
\beq
{\cal L}=Tr\left[i\psi^{\dagger}\partial_+\psi+i\chi^{\dagger}\partial_-\chi
-\frac{g^2}{2} J\frac{1}{\partial^2_-}J\right].
\eeq
It is obvious that the left-moving fields $\chi$ decouple, because 
their equations of motion do not involve a time derivative, {\em i.e.}~are
constraint equations.
Noting the simple expression of the interaction in terms of the
currents, it is natural to formulate the theory with $SU(N_c)$ currents 
as basic degrees of freedom.
For reasons of clarity, 
we will not use the terminology of bosonization or 
conformal field theory. We shall rather stick to the definition 
of the currents as a 
bilinear product of fermions, Eq.~(\ref{current}), and derive everything based on this definition,
which is perfectly possible.

The key issue is to obtain the mass
eigenvalues $M_n$ by solving the eigenvalue problem 
\beq\label{EVP}
{\cal M}^2|\varphi\rangle\equiv 2P^+P^-|\varphi\rangle=M^2_n|\varphi\rangle,
\eeq
where we act with the light-cone momentum and energy operators, $P^+$ and 
$P^-$, on a state $|\varphi\rangle$. 
The operators $P^\pm\equiv T^{+\pm}$
can be found by constructing the energy-stress tensor 
$T^{\mu\nu}$ in the canonical way, and one obtains 
\begin{eqnarray}
P^+&=&T^{++}=\frac{\pi}{N_c+N_f}\int^{\infty}_{-\infty} 
dx^- :Tr[J(x^-)J(x^-)]:\label{Pplus}\\
P^-&=&T^{+-}
=-\frac{g^2}{2}\int^{\infty}_{-\infty} 
dx^-:Tr[J(x^-)\frac{1}{\partial^2_-}J(x^-)]:.
\end{eqnarray}
The Sugawara form of the momentum operator $P^+$, 
Eq.~(\ref{Pplus}), might seem
somewhat unfamiliar, but an explicit analysis of this construction
in terms of the 
fermionic mode operators yields indeed the above result  
\footnote{The remainder of the product of four fermion operators
is the contraction term. Its integral becomes the momentum factor 
in the usual definition 
$P^+=\int_0^{\infty} dp\; p\; Tr[ b(-p)b(p)]$.}.
To solve the eigenvalue problem, Eq.~(\ref{EVP}), we have to diagonalize 
the mass squared operator, 
which is equivalent to diagonalizing the Hamiltonian $P^-$,
since $P^+$ is already diagonal. The latter is not as obvious as usual,
and we elaborate on this in Sec.~\ref{SecSpecial}. 

We use the standard mode expansion of the fermionic fields
\beq
\psi^{\, j}_k(x^-)=\frac{1}{2\sqrt{\pi}}\int_{-\infty}^{\infty}dp\, e^{-ipx^-} 
b^{\, j}_k(p),\label{psiExpansion}
\eeq
and the mode expansion of currents becomes
\begin{eqnarray}
J^{\, j}_k(p)&=&\frac{1}{\sqrt{2\pi}}\int_{-\infty}^{\infty}dx^- \,
e^{+ipx^-} J^{\, j}_k(x^-)
= \frac{1}{\sqrt{2\pi}}\int_{-\infty}^{\infty}dq :b^{\, j}_l(q)b^{\, l}_k(p-q):.
\label{current_mode}
\end{eqnarray}
The canonical anti-commutation relation for the fermionic operators
\begin{equation}
\{b^{\, j}_k(p),b^{\, l}_m(p')\}=\delta(p+p')\delta^{\, j}_m\delta^{\, l}_k,
\label{bbCR}
\end{equation}
determines the commutator of current with fermionic modes
\begin{equation}
[J^{\, j}_k(p),b^{\, l}_m(p')]=\delta^{\, l}_k b^{\, j}_m(p+p')
-\delta^{\, j}_m b^{\, l}_k(p+p').
\end{equation}
Recall that in the adjoint theory $b^{\, j}_k(-n)=b_j^{\dagger k}(n)$.
Due to the occurrence of Schwinger terms the current-current commutator is 
harder to derive. It is, however, well-known that the modes of the currents are
subject to the Kac-Moody algebra \footnote{We use the opportunity to 
correct Ref.~\cite{UT}, where the $\cal T$-symmetric 
({\em cf.}~Eq.~(\ref{Tsym}))
version of the algebra was used, with
no consequences for the results in the bosonic sector.}
\begin{equation}
[J^{\, j}_k(p),J^{\, l}_m(p')]=pN_f\delta^{\, j}_m\delta^{\, l}_k
\delta(p+p')+
\delta^{\, l}_k J^{\, j}_m(p+p')-\delta^{\, j}_mJ^{\, l}_k(k+k').
\label{KMCR}
\end{equation}
The vacuum is defined by
\begin{equation}
J^{\, j}_k(p)|0\rangle=0, \quad\mbox{ and } \quad 
b^{\, j}_k(p)|0\rangle=0,\; \quad \forall p\ge 0.
\label{vacuum}
\end{equation}
Following the usual DLCQ program \cite{PauliBrodsky85a}, 
we put the system in a box of length $2L$ and
impose anti-periodic boundary conditions on the fermionic fields 
$\psi(x^--L)=-\psi(x^-+L)$. 
The currents are by construction
subject to periodic boundary conditions.
The momentum modes are now discrete, and, as always 
in light-cone quantization, the longitudinal momenta are non-negative. 
The smallest momentum $k_{min}=P^+/2K$ is determined by
the harmonic resolution $K\equiv P^+L/\pi$, 
which controls the coarseness of the momentum-space discretization. 
The continuum limit is obtained by sending $K$ to infinity. 
In practice one solves the eigenvalue problem, Eq.~(\ref{EVP}), for growing values of $K$
and extrapolates the spectrum to the continuum by {\em e.g.} 
fitting the eigenvalues to a polynomial in $1/K$.
The expansion of the fermion fields,
Eq.~(\ref{psiExpansion}), becomes
\beq
\psi^j_k(x^-)=\frac{1}{2\sqrt{L}}\sum_{n=\pm\frac{1}{2},\pm\frac{3}{2},\ldots} 
B^j_k(n) e^{-i\pi nx^-/L}, 
\eeq
with the discrete field 
operators $B^j_k(n){\equiv}({\pi/L})^{1/2}b^{\, j}_k(n\pi/L)$.
The current mode operators $J(n)$ are defined by the discrete version 
of Eq.~(\ref{current_mode}).
The momentum operators read
\begin{eqnarray}
P^+&=&\left(\frac{\pi}{L}\right)
\frac{1}{N_c+N_f}Tr\left[\frac{1}{2}J(0)J(0)
+\sum^{\infty}_{n=1} J(-n)J(n)\right],\\
P^-&=&
\frac{\tilde{g}^2}{2\pi}\sum^{\infty}_{n=1} \frac{1}{n^2}Tr[J(-n)J(n)],
\end{eqnarray}
and become finite-dimensional matrices on the 
Hilbert space constructed by acting with the current operators
of momentum $K$ or smaller
on the vacuum defined by Eq.~(\ref{vacuum}). 
For convenience we introduced 
the scaled coupling $\tilde{g}^2\equiv {g^2L}{/\pi}$. 
We emphasize the appearance of the zero mode contribution 
in the discrete formulation, as should be clear from
$P^+=\lim_{\epsilon\rightarrow 0}\frac{1}{N_c+N_f}Tr\left[
\frac{\epsilon}{2}J(0)J(0)+
\int_{\epsilon}^{\infty}dp J(-p)J(p)\right]$.
In the Veneziano limit the operators are realized on a 
Hilbert space of discrete $SU(N_c)$ singlet Fock states.
The fermionic states look like
\begin{equation}
|{\bf b}+\frac{1}{2}; n_1,\ldots,n_b\rangle=
(N_c N_f)^{-b/2-1/4}
Tr[J(-n_1)J(-n_2)\cdots J(-n_b)B(-\frac{1}{2})]|0\rangle,\label{fState}
\end{equation}
whereas in the bosonic sectors we find the singlets 
\begin{equation}
|{\bf b}; n_1,\ldots,n_b\rangle=(N_c N_f)^{-b/2}
Tr[J(-n_1)J(-n_2)\cdots J(-n_b)]|0\rangle.
\label{bState}
\end{equation}
The additional fermion operator $B^{\, j}_k(-1/2)$ in the fermionic 
states is the source of some rather odd differences between the two sectors, 
as we shall see in the next section.

\section{Specialties of the fermionic sector}
\label{SecSpecial}

The numerical 
solution to the eigenvalue problem, Eq.~(\ref{EVP}), in the bosonic 
sector was presented 
in Ref.~\cite{UT}. The calculations in the fermionic sector
are not quite as straightforward, and we shall point out the major 
differences.
The first peculiarity in the fermionic sector
resides in the action of the momentum operator 
$P^+$ on a fermionic state. It turns out that
the eigenvalue of $P^+$ on a fermionic state depends on the ratio $\lambda=N_f/N_c$. 
To convince ourselves that this is true, we consider the 
simplest case in discrete formulation
\begin{eqnarray}
P^+Tr\left[J(-n)B\left(-\frac{1}{2}\right)\right]|0\rangle\!\!&{=}&\!\!
\left[P^+,J^i_j(-n)\right]B^{\, j}_i\left(-\frac{1}{2}\right)|0\rangle
+J^i_j(-n)\left[P^+,B^{\, j}_i\left(-\frac{1}{2}\right)\right]|0\rangle\nonumber\\
&&+\left[\left[P^+,J^i_j(-n)\right],
B^i_j\left(-\frac{1}{2}\right)\right]|0\rangle\nonumber\\
&=&\frac{\pi}{L}
\left(n+\frac{1}{1+\lambda}\right)Tr\left[J(-n)B\left(-\frac{1}{2}\right)\right]|0\rangle.
\end{eqnarray}
In other words, only in the adjoint theory the fermion has the familiar 
momentum. 
In the 't Hooft limit, a fermion with half-integer momentum 
contributes the same momentum as a current, whereas in the 
large $N_f$ limit it has a vanishing contribution.
This is a consequence of the discrete formulation.
In particular, it is 
${\cal M}^2 B^i_j(-1/2)|0\rangle=2P^+P^-B^i_j(-1/2)|0\rangle=0$, 
although $P^+B^i_j(-1/2)|0\rangle=(1+\lambda)^{-1}B^i_j(-1/2)|0\rangle$.

Another issue to be addressed here is the size of the Fock space.
In the bosonic sector, the singlet states are 
of the form of Eq.~(\ref{bState}), {\em i.e.}~they are single-trace
states of a certain number of currents. This form allows for cyclic 
permutations of the currents under the trace. The cyclic permutations
are related non-trivially due to the Kac-Moody structure of the currents,
yet all cyclic permutations of a given state have to be eliminated from the
Fock basis.
The key difference in the fermionic sector is the absence
of these cyclic permutations; 
the fermion defines the ordering of the state.
This is quite natural, since it basically acts like an adjoint 
vacuum, as we shall see. This renders the Fock 
basis much larger than in the bosonic case.
The number of states grows like $2^{K-3/2}$, see Table \ref{TableSize},
with the harmonic resolution $K$ being a half-integer in the fermionic 
sector due to the momentum of the additional fermion. 
The different sizes of the Fock bases
will help us interpreting the resulting fermionic spectra, 
because the spectra in the 't Hooft and large $N_f$ limits of the theory 
have the same single-particle content as in the bosonic sectors.


We briefly comment on the 
fact that we can calculate two eigenvalues of the mass squared operator
${\cal M}^2=2P^+P^-$ analytically.
Remarkably, we obtain the same functional form $M^2_{1,2}(K)$
as in the bosonic sector.
This is somewhat surprising, since in the fermionic sector we loose the
uniqueness of the two states with the largest number of currents.
At harmonic resolution $K=b+1/2$ the states 
\begin{eqnarray}
|b+\frac{1}{2}\rangle&=&Tr\left[\{J(-1)\}^{b}B\left(-\frac{1}{2}\right)
\right]|0\rangle
\label{K}\\
|b-\frac{1}{2}\rangle&=&
Tr \left[\{J(-1)\}^{b-2} J(-2)B\left(-\frac{1}{2}\right)\right]|0\rangle,\label{K-1}
\end{eqnarray}
have $b$ and $b-1$ currents, respectively, plus a fermion with momentum 
$1/2$. However, Eq.~(\ref{K-1}) represents a class of $b-1$ states,
rather than a unique state as in the bosonic sector.
Explicit calculations show that the image of one of 
these states under the mass squared operator 
has no overlap with any of the other states, and its eigenvalue 
can be trivially extracted in the fashion of Ref.~\cite{UT}. Hence,
two eigenvalues can be evaluated {\em a priori}. 
The eigenvalues of the mass squared operator associated with the states,
Eq.~(\ref{K}) and (\ref{K-1}), are 
\begin{eqnarray}\label{analyticEVs}
M^2_1(K=b+\frac{1}{2})&=&\frac{g^2 N_c}{\pi}(1+\lambda)\left(b+\frac{1}{1+\lambda}\right)^2,\\
M^2_2(K=b+\frac{1}{2})&=&\frac{g^2 N_c}{\pi}(1+\lambda)\left(b+\frac{1}{1+\lambda}\right)
\left(b-\frac{3}{2}+\frac{1}{1+\lambda}\right).
\end{eqnarray}
We note that in the 't Hooft limit, all states 
of the form Eq.~(\ref{K-1}), have the same eigenvalue.


\section{The Hamiltonian}
\label{SecHamiltonian}

We construct the light-cone Hamiltonian in the framework of DLCQ.
Once the commutation relations, Eq.~(\ref{bbCR})--(\ref{KMCR}), 
are specified, 
and the Fock basis is chosen,
this is a straightforward generalization of previous work \cite{UT}, 
and we can be brief here.
In the fermionic sector we get additional
contributions to the light-cone Hamiltonian by commuting through zero modes
of the currents and acting with them on the extra fermion.
Note that annihilation operators may be created by commuting current operators.
To streamline the calculations it is useful to distinguish creation,
annihilation and zero-mode part of a commutator
\[
[A,B]\equiv \lceil A,B\rceil + \lfloor A,B\rfloor+ \lfloor A,B\rfloor_0,
\]
much in the fashion of Ref.~\cite{UT}.
The resulting Hamiltonian is slightly simpler 
than in the bosonic case.
The number of its terms of leading power 
in $N_c$ grows exactly quadratic with the number $b$ 
of currents in a state.
In the large $N_c$ limit the action of $P^-$ on a state 
with $b$ currents, Eq.~(\ref{fState}), is then
\begin{eqnarray}\label{Pminus}
&&\!\!\!\!\!\!\!\!\!\!\!\!\!\!P^-|{\bf b}+\frac{1}{2}; n_1,\ldots,n_b\rangle
\nonumber\\
%
%
&=&\frac{\tilde{g}^2N_c}{2\pi}\sum_{i=1}^b
\left(\sum_{m=1}^{n_i-1}\frac{1}{(m-n_i)^2}-
\sum_{m=1}^{n_i}\frac{1}{m^2}\right)
|{\bf b+1}+\frac{1}{2};n_1,n_2,\ldots,n_i-m,m,\ldots, n_b,
\frac{1}{2}\rangle\nonumber\\
%
%
&&+\frac{\tilde{g}^2N_c}{2\pi}\sum_{i=1}^b\left(\frac{\lambda}{n_i}
+\sum_{m=1}^{n_i-1}\frac{1}{m^2}\right)
|{\bf b}+\frac{1}{2};n_1,n_2,\ldots, n_b,\frac{1}{2}\rangle\nonumber\\
%
%
&&+\frac{\tilde{g}^2N_c}{2\pi}\sum_{i=1}^{b-1}
\left(\sum_{m=0}^{n_i-1}\frac{1}{(m+n_i)^2}
-\sum_{m=1}^{n_i-1}\frac{1}{m^2}\right)
|{\bf b}+\frac{1}{2};n_1,\ldots, n_i+m,n_{i+1}-m, 
\ldots n_b,\frac{1}{2}\rangle\nonumber\\
%
%
&&+\lambda\frac{\tilde{g}^2 N_c}{2\pi}\sum_{j=1}^{b-1}
\sum_{i=1}^{b-j}\left[
\frac{1}{(\sum_{q=i}^{j+i}n_q)^{2}}-
\frac{1}{(\sum_{q=i+1}^{j+i}n_q)^{2}}
\right]n_{i+j}\nonumber\\
&&\quad\quad\quad\quad\times
|{\bf b-j}+\frac{1}{2};n_1,n_2,\ldots,n_{i-1},\sum_{q=i}^{j+i}n_q,n_{j+i+1},
\ldots, n_b,\frac{1}{2}\rangle\nonumber\\
%
&&+\frac{\tilde{g}^2N_c}{2\pi}\sum_{j=1}^{b-2}
%
\sum_{i=1}^{b-j-1}\sum_{m=0}^{n_{i+j+1}-1}\left(
\frac{1}{(m+\sum_{q=i}^{i+j}n_q)^2}-
\frac{1}{(m+\sum_{q=i+1}^{i+j}n_q)^2}\right)\nonumber\\
&&\quad\quad\quad\quad\quad\times
|{\bf b-j+\frac{1}{2}};n_1,n_2,\ldots,\sum^{i+j}_{q=i}n_{q}+m,n_{i+j+1}-m,
n_{i+j+2},\ldots, n_b,\frac{1}{2}\rangle\nonumber\\
%
%
&&+\frac{\tilde{g}^2 N_c}{2\pi}\sum_{i=1}^{b-1}
\left[
\frac{1}{(\sum_{q=b-i}^{b}n_q)^{2}}-
\frac{1}{(\sum_{q=b-i+1}^{b}n_q)^{2}}\right]
|{\bf b-i}+\frac{1}{2};n_1,\ldots,n_{b-i-1},\sum_{q=b-i}^{b}n_q,
\frac{1}{2}\rangle\nonumber\\
%
%
&&+\frac{\tilde{g}^2 N_c}{2\pi}
\left(
\frac{1}{n_1^2}+\frac{1}{n_b^2}\right)
|{\bf b}+\frac{1}{2};n_1,n_2,\ldots,n_{b},\frac{1}{2}\rangle.
\end{eqnarray}
As in the bosonic case \cite{UT}, the terms in the Hamiltonian have a 
different $N_c$ and $N_f$ behavior.
Only the terms in lines two and four of Eq.~(\ref{Pminus})
contain $N_f$. These terms will be
absent in the 't Hooft limit and will be dominant
in the large $N_f$ limit.

\begin{table}
\centerline{
\begin{tabular}{|c|cccccccccccccc|}\hline
$b$ &1 &2 & 3 & 4 & 5 &6 & 7& 8 & 9 & 10 & 11 & 12 & 13 &14\\\hline
fermions & 1 &2 & 4 & 8 &16& 32 &64 &128 &256 & 512 & 1024 & 2048 & 4096 &
8192\\
bosons & --- & 1 &2 & 4 & 6 &12& 18 &34 &58 &106 & 186 & 350 & 630 & 1180\\
\hline
\end{tabular}
}
\caption{Number of basis states a a function of the number $b$ of currents
in a state. The relation of $b$ to the harmonic resolution $K$ is $K=b$ 
in the bosonic, and $K=b+\frac{1}{2}$ in the fermionic sector.}
\label{TableSize}
\end{table}


\section{Numerical Results}
\label{SecNumCalc}

We solve 
the eigenvalue problem, Eq.~(\ref{EVP}), numerically 
to obtain the mass spectrum of the theory as a function of the harmonic 
resolution $K$ and the ratio of the number of colors and flavors, $\lambda$. 
Remember that the Veneziano limit is always understood.
In the sequel, we will use both 
parameters to extract information from the spectra. 
We recover the large $N_f$ limit when 
$\lambda\rightarrow \infty$, the 't Hooft 
model in the limit $\lambda\rightarrow 0$,
and the adjoint model, of chief interest, at $\lambda=1$.
It turns out that the complexity of the spectra grows in this
order, and we will start their interpretation with the rather simple 
large $N_f$ limit.
Since we are using a discrete formulation, we may aim to understand 
{\em all} states in the spectra.

Some words on the numerical algorithm seem in order.
The number of states grows exponentially with the harmonic 
resolution, {\em cf.}~Table \ref{TableSize}, and much faster than in the 
bosonic sector. However, the situation is still better than in the 
formulation of the theory with fermionic operators \cite{GHK97}; 
we have at $K=25/2$
a roughly four times smaller basis.
To further reduce the computational effort, we could use the 
$Z_2$ symmetry of the Hamiltonian which is invariant under the 
transformation
\begin{equation}
{\cal T}J_{ij}(n)=-J_{ij}(n).\label{Tsym}
\end{equation}   
It is straightforward to convince oneself that the action of this
operator on a state with $b$ currents is
\begin{equation}
{\cal T}|{\bf b}+\frac{1}{2}; n_1,n_2,\ldots,n_b\rangle=
\sum_{i=0}^{2^{b-1}}(-)^{p_i+1}|{\bf {p_i+1/2}}; 
\tau_i(n_b,n_{b-1},\ldots, n_1),\frac{1}{2}\rangle,\label{Tsym2}
\end{equation}
where the $\tau_i$ consist of ${p}_i$ partial sums of 
the $b$ momenta, in the sense that
$i$ runs over all possibilities to place $0,1,\ldots, b-1$ commas
between the 
momenta while summing those momenta which are not
separated by a comma, {\em e.g.} 
\begin{eqnarray}
{\cal T}|{\bf 7/2}; n_1,n_2,n_3,\frac{1}{2}\rangle
&=&|{\bf 7/2}; n_3,n_2,n_1,\frac{1}{2}\rangle
-|{\bf 5/2}; n_3,n_2+n_1,\frac{1}{2}\rangle\nonumber \\
&&-|{\bf 5/2}; n_3+n_2,n_1,\frac{1}{2}\rangle
+|{\bf 3/2}; n_3+n_2+n_1,\frac{1}{2}\rangle.
\end{eqnarray}
Since we do not work in an orthogonal basis, it is not very
helpful to block diagonalize the Hamiltonian with respect to this symmetry.
We will rather determine the $Z_2$ parity of an eigenstate {\em a posteriori} 
by calculating the expectation value of the operator ${\cal T}$ 
in this state. 
The main benefit of the {\em a priori} symmetrization, namely the 
reduction of the numerical effort to diagonalize a smaller matrix,
is of course lost this way. However,
the separation of the $Z_2$ odd and even eigenfunctions is very useful when
interpreting the results, because it reduces the density of eigenvalues to
roughly a half.

\begin{figure}
\centerline{
\psfig{figure=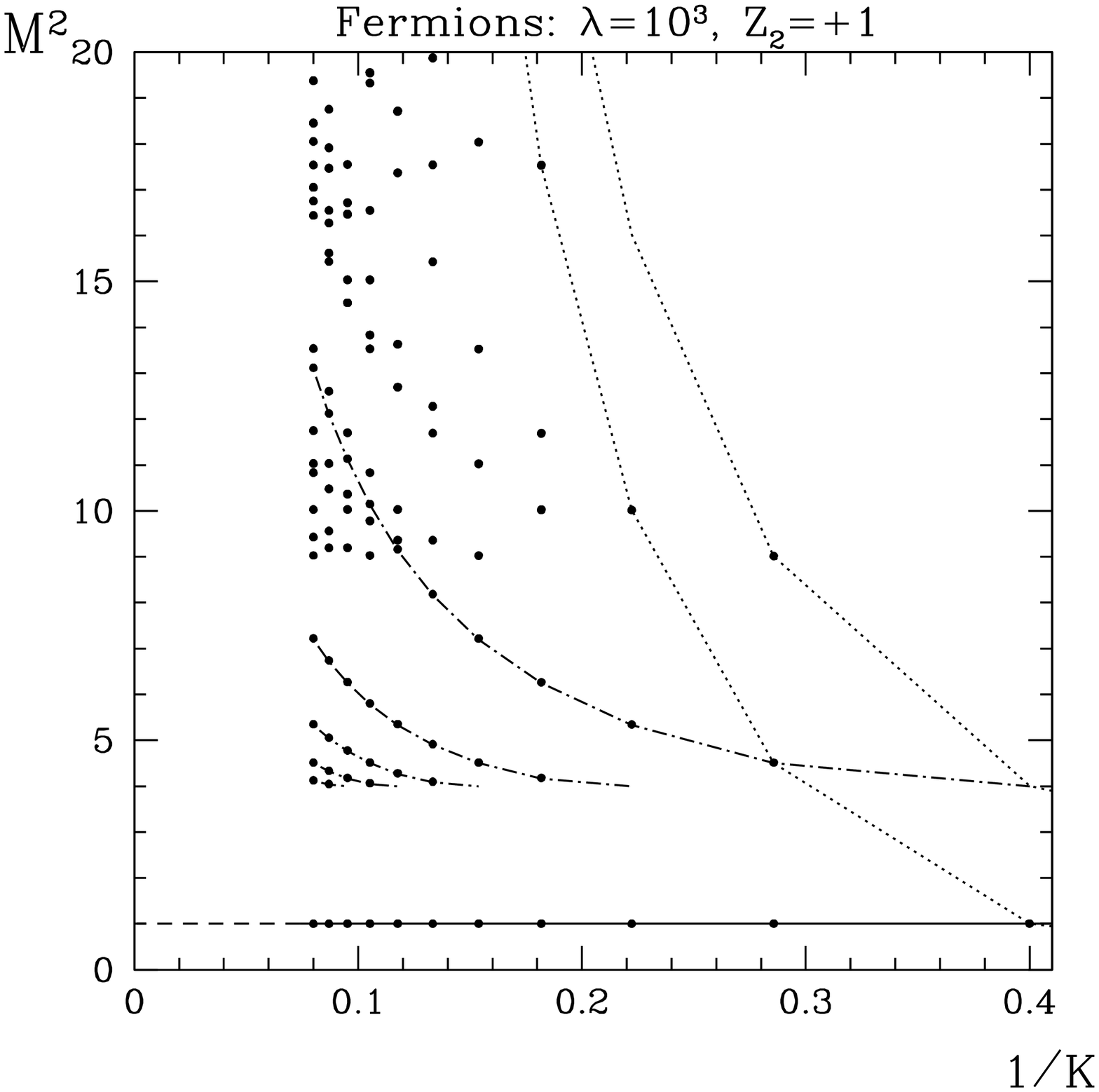,width=8cm,angle=0}
\psfig{figure=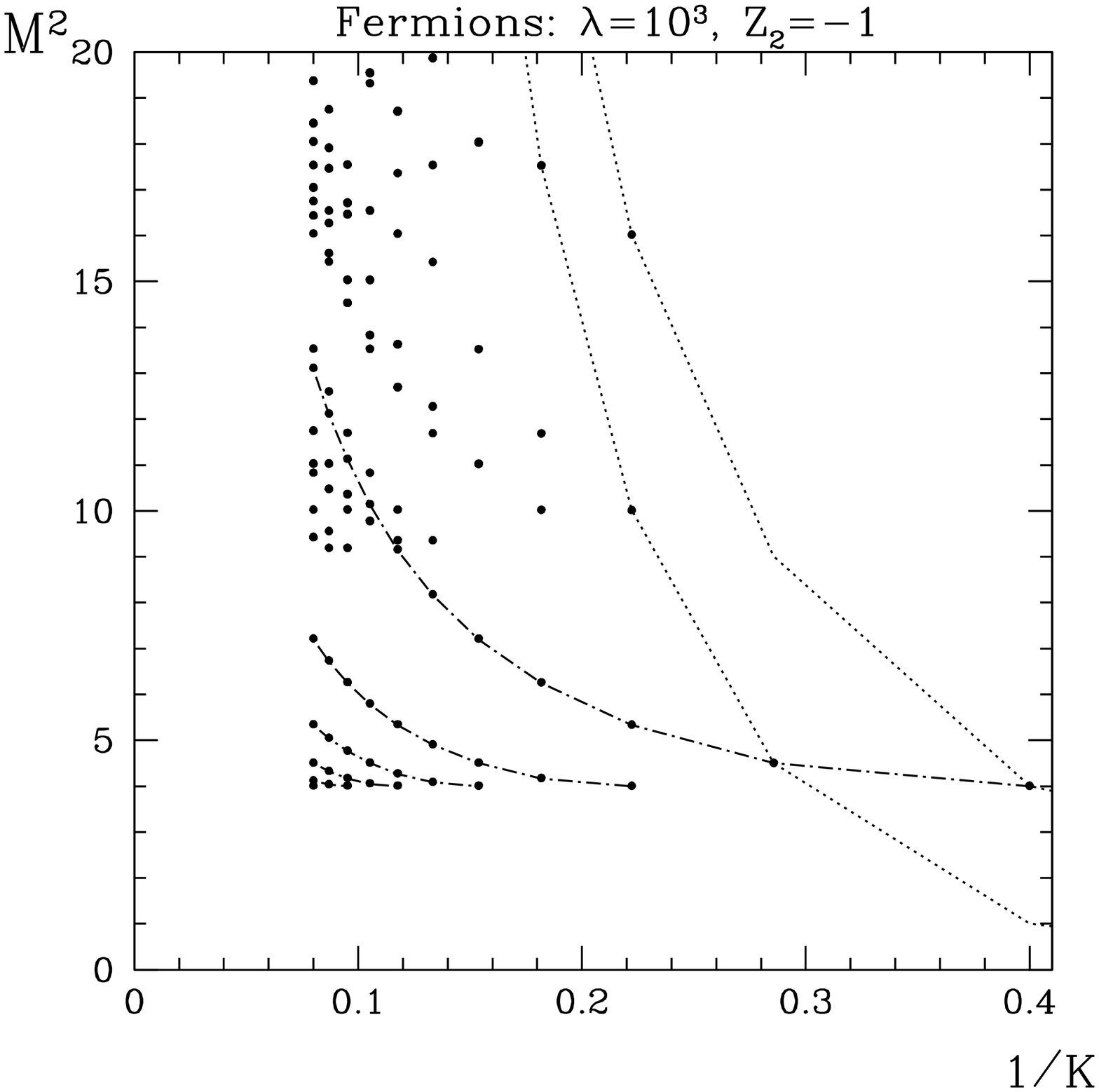,width=8cm,angle=0}}
\caption{Fermionic spectrum in the large $N_f$ limit.
Left: (a) $Z_2$ even sector. Right: (b) $Z_2$ odd sector.
Solid (dash-dotted) lines connect associated single(multi)-particle 
eigenvalues at different $K$.
Dotted lines connect analytically calculable eigenvalues. Dashed 
lines are extrapolations to the continuum limit.
Note that masses are in units $g^2N_f/\pi$.}
\label{largeNfSpectrum}
\end{figure}

\subsection{The large $N_f$ limit}
\label{SecLargeNf}

In the large $N_f$ limit 
we find the expected meson \cite{AFST98} in the {fermionic} $Z_2$ even sector
at mass
\begin{equation}
M^2_M(K)\equiv \frac{g^2N_f}{\pi},\label{MesonMass}
\end{equation} 
{\em cf.~}Fig.~\ref{largeNfSpectrum}.
All other states are multi-particle states built from this state, 
and are in this sense trivial.
It is, nevertheless, important that we understand {\em all} states in the
spectrum to see what we can learn in order to decode the adjoint spectrum.
Let us first focus on the fermionic sectors. 
It is easy to write down a formula for the mass of a 
multi-particle state consisting of non-interacting partons in DLCQ.
If we take into account the finding of Sec.~\ref{SecSpecial} that the momentum
of a fermion depends on $\lambda$, it reads
\begin{eqnarray}
M^2_{p_1,p_2,\ldots,p_{b}}(n_1,n_2,\ldots, n_{b-1};K)=
\left(K-\frac{1}{2}+\frac{1}{1+\lambda}\right)\hspace*{5cm}\nonumber\\ 
\times
\left(
\frac{M^2_{p_b}(K-\sum_{i=1}^{b-1}n_i+(b-1)[\frac{1}{2}-(1-\lambda)^{-1}])}
{K-\sum_{i=1}^{b-1}n_i+(b-2)[\frac{1}{2}-(1-\lambda)^{-1}]}
+\sum_{i=1}^{b-1}\frac{M^2_{p_i}(n_i)}{n_i-\frac{1}{2}+(1+\lambda)^{-1}}
\right).
\label{MultiFermions}
\end{eqnarray}
The masses of the multi-particle states grow like the momentum cutoff $K$, 
{\em i.e.} diverge in the continuum limit. We found it therefore,
contrary to previous work \cite{GHK97,UT}, 
appropriate to connect the multi-particle states reflecting this fact
in Fig.~\ref{largeNfSpectrum}.
Incidentally, this makes the labeling of states easier.
A multi-particle state with $b$ constituents is characterized by $b-1$
momenta $n_i$ and $b$ numbers 
$p_i$ specifying its single-particle constituents.
Eq.~(\ref{MultiFermions}) is slightly more general than needed here 
and holds also in the 
't Hooft limit. 
In the large $N_f$ limit there is only one single-particle state and it has
a constant mass, Eq.~(\ref{MesonMass}).
If we denote it by the operator $A^\dagger_M(n)$,
the Fock basis in the fermionic sector looks like 
\begin{eqnarray}
|1\rangle_F&=&A^{\dagger}_M\left(K-\frac{1}{2}\right)B^{\dagger}\left(\frac{1}{2}
\right)|0\rangle
\label{NfMeson}\\\
|2;n\rangle_F&=&A^{\dagger}_M\left(n\right)A^{\dagger}_M\left(K-\frac{1}{2}-n\right)
B^{\dagger}\left(\frac{1}{2}\right)
|0\rangle\label{Nf2Mesons}\\
|3;n,m\rangle_F&=&A^{\dagger}_M\left(n\right)A^{\dagger}_M\left(m\right)
A^{\dagger}_M\left(K-\frac{1}{2}-n-m\right)B^{\dagger}\left(\frac{1}{2}\right)|
0\rangle, 
\mbox{ etc.}\label{FockSolution}
\end{eqnarray}
Fock basis states with $b$ mesons 
will thus be constructed by assigning meson 
momenta as all partitions of $K-1/2$ into 
$b$ integers, {\em i.e.}~exactly like the states, Eq.~(\ref{fState}), 
except that now we are operating with meson rather than current operators. 
The role of the fermion in the states will be discussed in
Sec.~\ref{SecAdjoint}. 
Here it serves as a convenient tool for book-keeping.
The assignment of quantum numbers $\pi_{\cal T}$ of the $Z_2$ symmetry, 
Eq.~(\ref{Tsym}), is clear:
the single particle state Eq.~(\ref{NfMeson}) 
is a two-parton state and therefore according 
to Eq.~(\ref{Tsym2}) $Z_2$ even, $\pi_{\cal T}=+1$. 
It comprises a boson and a fermion with unique momentum partition and we
find it indeed only in the spectrum of the fermionic $Z_2$ even sector.
The states with two mesons, Eq.(\ref{Nf2Mesons}), 
are actually three parton states
in the fermionic sector. Therefore the states where the mesons have the
same momentum transform to minus themselves under the $Z_2$ symmetry,
and are present only in the $Z_2$ odd sector. 
All other momentum partitions should be present in both $Z_2$ sectors.
This is exactly what we see in the spectra.
The generalization to the $b$ parton states is obvious,
and reproduces exactly the spectra in Fig.~\ref{largeNfSpectrum}.


An analogous construction can be used in the bosonic sectors.
The spectra are depicted in Fig.~5(a) of Ref.~\cite{UT}. 
Again we build up the Fock basis from solutions of the 
dynamical eigenvalue problem, Eq.~(\ref{EVP}). But now the 
meson operators will act on the vacuum itself, rather than 
on the fermion and the vacuum
\begin{eqnarray}
|1;n\rangle_B&=&A^{\dagger}_M(n)A^{\dagger}_M(K-n)|0\rangle
\label{BNfMeson}\\\
|2;n,m\rangle_B&=&A^{\dagger}_M(n)A^{\dagger}_M(m)A_M^{\dagger}(K-n-m)
|0\rangle, \quad\mbox{etc.}
\end{eqnarray}
Consequently, we have to discard all states with momentum partitions
equivalent under cyclic permutations. This is, of course, nothing else than
constructing the current basis in the bosonic sectors, as we did in 
Ref.~\cite{UT}. 
A two-meson state is a two-parton state in the bosonic sector. 
Hence, the $Z_2$ quantum numbers are opposite as in the fermionic sector.
In particular, since there are no cyclic permutations of momenta of 
the currents, the two-meson states are absent altogether in the 
in the bosonic $Z_2$ odd sector, as observed. The generalization
to $b$ partons is again obvious, 
and we thus {completely} constructed the spectrum of 
the large $N_f$ limit of the theory.
\begin{figure}
\centerline{\psfig{figure=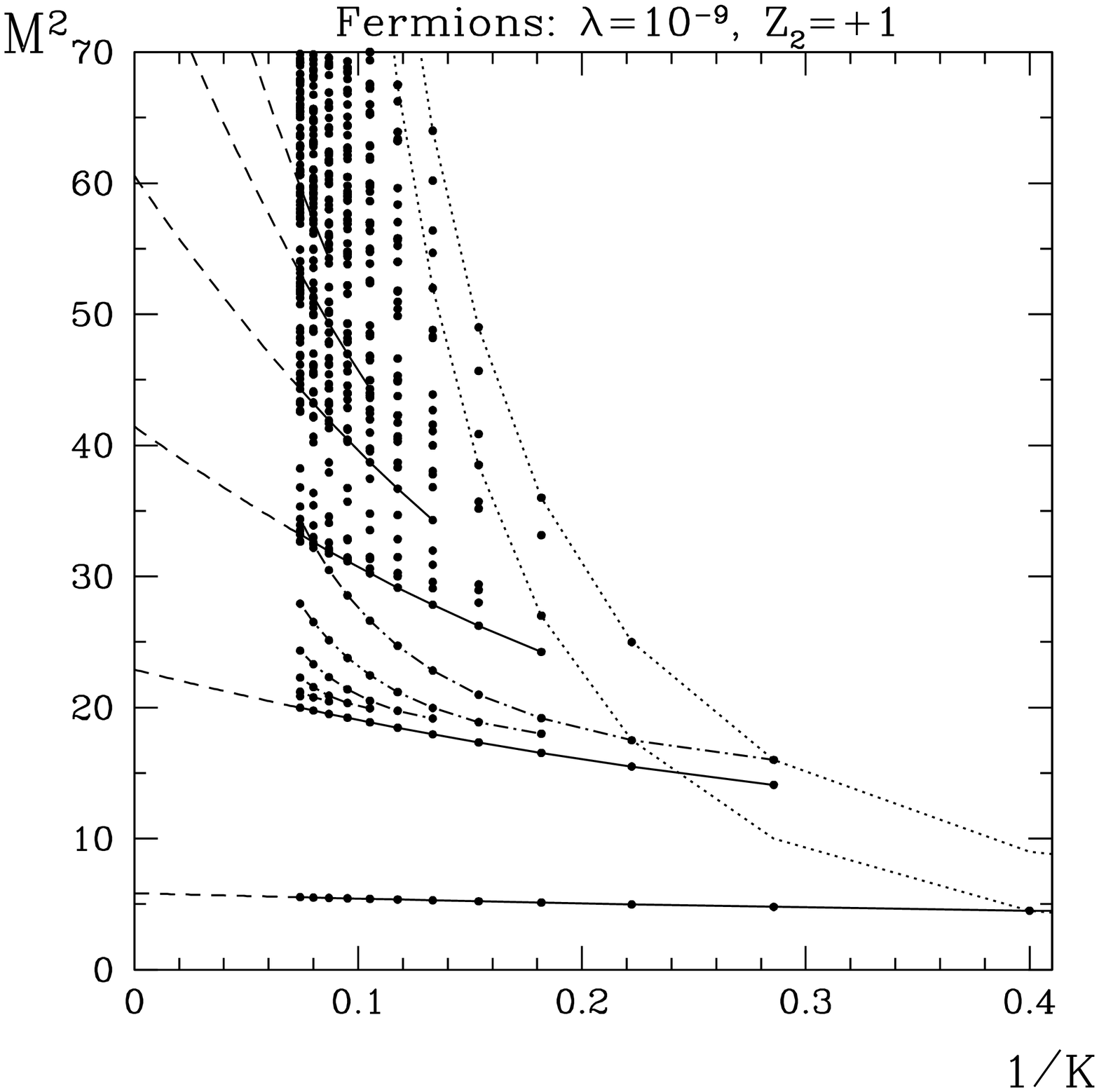,width=8cm,angle=0}
\psfig{figure=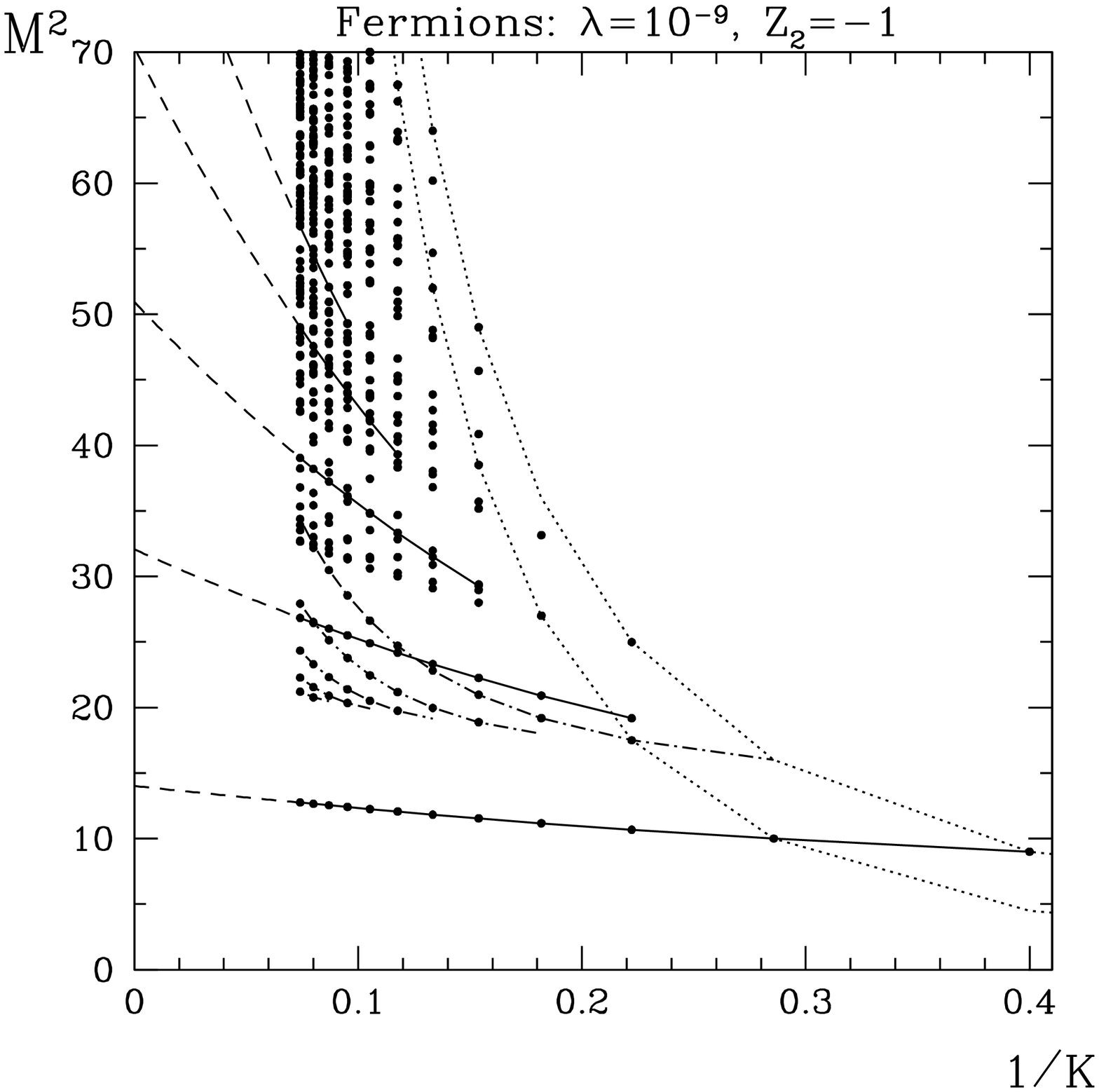,width=8cm,angle=0}
}
\caption{Fermionic spectra of the 't Hooft limit in the $Z_2$ even (a) and
odd (b) sectors. 
Solid (dash-dotted) lines connect associated single(multi)-particle 
eigenvalues at different $K$.
Dotted lines connect analytically calculable eigenvalues. Dashed 
lines  are extrapolations to the continuum limit. Masses are in units 
$g^2N_c/\pi$.}
\label{tHooftSpectrum}
\end{figure}

\subsection{The 't Hooft limit}
\label{SecTHooft}

In the 't Hooft limit
we find {exactly} the same eigenvalues as in previous work \cite{UT}, 
Eq.(29), namely
\beq
M^2= 5.88,14.11,23.04,32.27,41.68,51.24,60.93,70.76,80.97,90.90,
\eeq
{\em cf.~}Fig.~\ref{tHooftSpectrum},
which is in very good agreement with 't Hooft's original solution 
\cite{tHooft74b}.
These continuum results are obtained 
by fitting the eigenvalue trajectories $M_i^2(K)$ of the single-particle 
states to polynomials of second
order in $1/K$ and subsequently extra\-polating to $K\rightarrow\infty$.
If we compare the spectrum, Fig.~\ref{tHooftSpectrum}, to the bosonic sector,
Fig.~2 of Ref.~\cite{UT}, we find much more multi-particle 
states and, of course, a shift of half a unit in momentum. 
As in the large $N_f$ limit, the multi-particle states decouple, but now we 
have several single-particle states. Namely, the spectrum at resolution
$K$ contains $K-1/2$ 't Hooft mesons. The $i$th meson makes its first appearance
at resolution $K=i+1/2$ and has $\pi_{\cal T}=(-1)^{i+1}$.  
Note that the single-particle masses are functions of $K$.
If we would set up an orthonormal current basis for this problem to factorize
the Hamiltonian into its single- and multi-particle blocks,  
the task would be to diagonalize a $(K-1/2)\times (K-1/2)$ 
matrix to find
the masses of these mesons. Instead we are diagonalizing a $2^{K-3/2}$ 
dimensional matrix. The first procedure
is, however, due to the tedious evaluation of the scalar product
of Kac-Moody states more expensive than to actually diagonalize the much larger
matrix \cite{UT}. 


We already wrote down the formula for the masses of the multi-particle 
states, Eq.~(\ref{MultiFermions}). Also the designation of $Z_2$
quantum numbers from the partitions of the parton momenta stays the same
as in the large $N_f$ limit.
It should be noted, however, that due to the difference in effective momenta
of the fermion in the states, the first multi-particle 't Hooft state appears
at $K=7/2$, as opposed to $K=5/2$ in the large $N_f$ limit. 
As in the large $N_f$ limit, 
we reproduce the distribution of the $Z_2$ even and odd states, 
and understand the spectra completely in terms of their 
single-particle content.
In particular, there is no need for recurring
to properties of the massless sector
of the theory, because 
we can construct all quantum numbers from the information
of the massive spectrum. This will change substantially in the adjoint case
which we consider next.

\subsection{Adjoint fermions}
\label{SecAdjoint}

Solving the eigenvalue problem, Eq.~(\ref{EVP}) in the adjoint model, $\lambda=1$,
we obtain precisely the same eigenvalues as previous 
works with fermions as basic degrees of freedom, 
{\em e.g.} Ref.~\cite{GHK97}, 
with {\em anti-periodic} boundary 
conditions for the fermions. This is not surprising because the 
formulation of the theory with $SU(N_c)$ currents rather than with fermions
is in essence a change of basis.

\begin{figure}
\centerline{
\psfig{figure=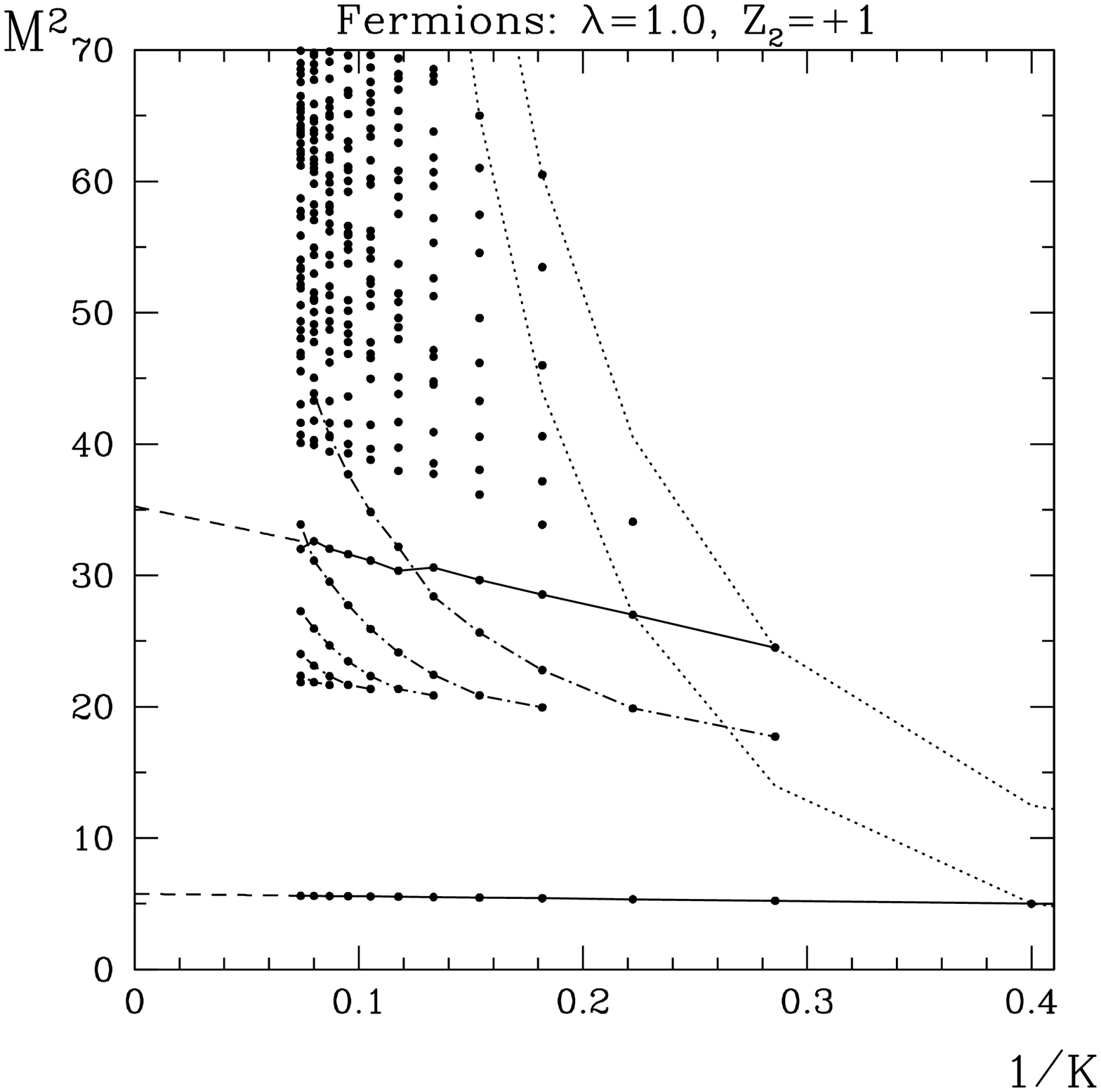,width=8cm,angle=0}
\psfig{figure=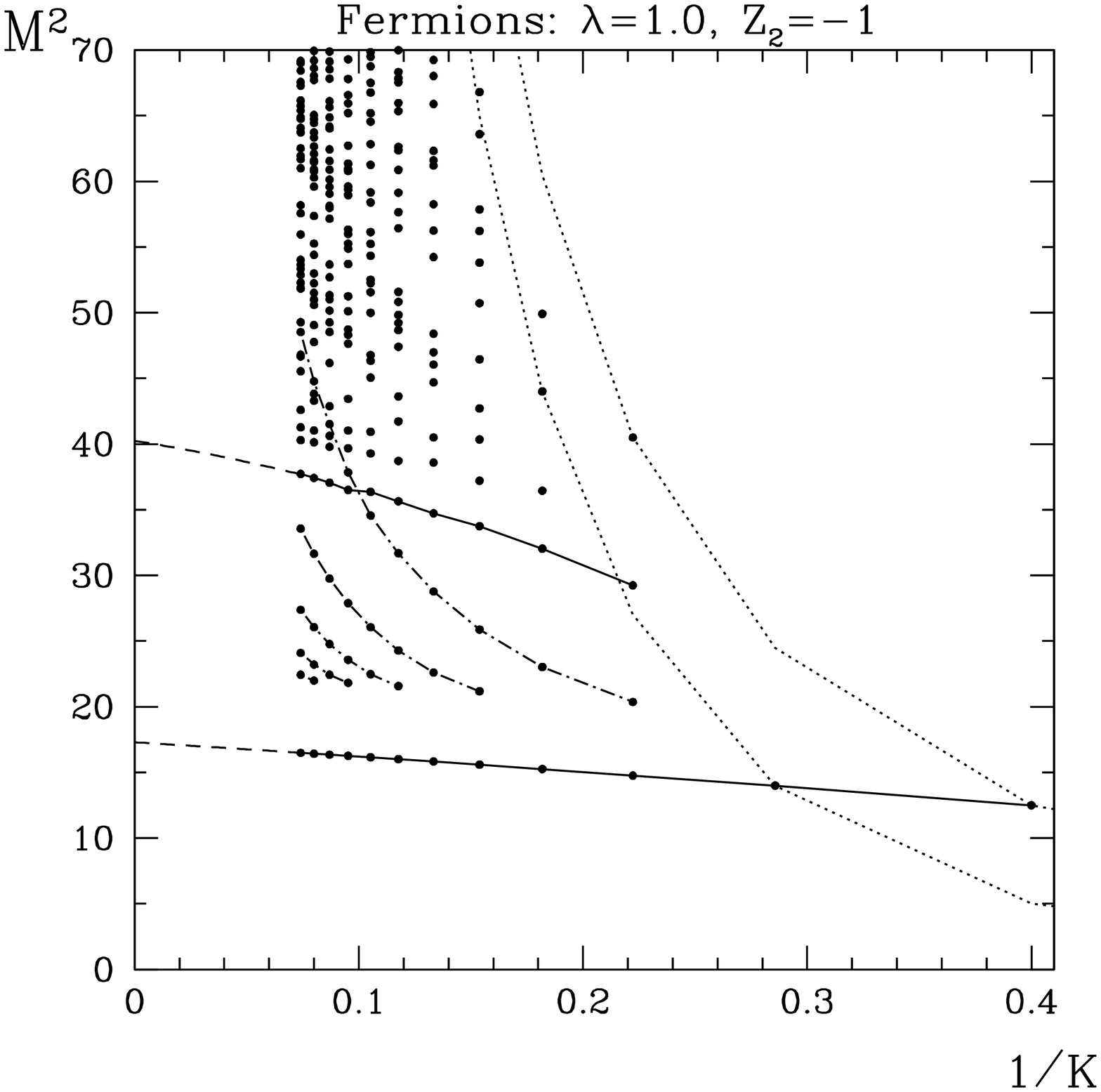,width=8cm,angle=0}
}
\caption{Fermionic spectra of the theory with 
adjoint fermions in the $Z_2$ even (a) and
odd (b) sectors. 
Solid (dash-dotted) lines connect conjectured single(multi)-particle 
eigenvalues at different $K$.
Dotted lines connect analytically calculable eigenvalues. Dashed 
lines are extrapolations to the continuum limit.
Masses are in units $g^2N_c/\pi$.}
\label{AdjointSpectrum}
\end{figure}

If we look at the eigenvalue trajectories (mass squared as a function of $K$)
in Fig.~\ref{AdjointSpectrum},
the structure of the spectrum looks similar to the \mbox{'t Hooft} case.
We see immediately four single-particle candidates 
which qualify by their quasi-linear trajectories. 
In the continuum limit they have the eigenvalues
\beq
M^2_{F_1}=5.75,\quad M^2_{F_2}=17.29, \quad M^2_{F_3}=35.25, 
\quad M^2_{F_3}=40.24,
\eeq
in units ${g^2N_c}/{\pi}$, see also Table \ref{Table2}.
It is, however, questionable if these states are indeed single-particle states.
There are a couple of problems which prevent a straightforward interpretation
of the adjoint spectrum,
which will become clear when we compare the adjoint to the 't Hooft spectrum.
In the adjoint spectrum we find kinks in the single-particle trajectories,
and also the multi-particle trajectories are distorted. This is clear 
evidence for an interaction between these states. Since the multi-particle
states do not decouple, a mass formula analogous to Eq.~(\ref{MultiFermions}) 
cannot be exact.
Furthermore, the masses of the 
single-particle states of the fermionic and bosonic
sectors are not degenerate as in the 't Hooft case,
but differ significantly. This in turn means that we cannot have a
fermionic and bosonic Regge trajectory of single-particle states, if all
of them give rise to multi-particle states: there would be simply too many states 
to account for in a discrete Fock basis.

As a way out of this dilemma, we make the following conjecture which we will
try to support by empirical facts.
Namely, we view the sole fermion in the states of smallest discrete 
momentum $k_{min}=P^+/2K$ acting on the vacuum 
as the finite $K$ expression for an ``adjoint vacuum'',
as it appears in the bosonized version of this theory
\cite{ArmoniSonnenschein95}. It is clear that the 
correct expression should involve a fermionic zero mode, which is 
absent in the present discrete approach. It is recovered
in the continuum limit $K\rightarrow\infty$.
This conjecture makes the following interpretation of the 
spectrum plausible. 
The approximate vacuum 
will introduce couplings between states that are decoupled in the 
continuum limit. 
Most of these states will be multi-particle states, which are
necessary to describe the theory correctly at finite 
harmonic resolution $K$. Since the 
approximation occurs only in the fermionic sector, it seems natural that 
the artifacts associated with this finite $K$ effect also 
originate in this sector. 
In other words, 
we expect {\em only} the fermionic single-particle states 
to give rise to multi-particle states.
We can check this conjecture by looking at the spectra at small 
$\lambda$. There we expect that the multi-particle states are still
very well described by the DLCQ formula for the spectrum of free 
particles, Eq.~(\ref{MultiFermions}). On the other hand, 
the masses of the lowest single-particle boson and fermion are 
already noticeably different. By constructing two sets of multi-particle 
states out of two bosons and two fermions, respectively, we convinced 
ourselves that the eigenvalues are well described by a pair of free
fermions, but not of bosons.
On the other hand, we expect the coupling between the single- and 
multi-particle states 
to vanish as the harmonic resolution $K$ grows.
In particular, the deviations from a multi-particle mass formula
analogous to Eq.~(\ref{MultiFermions}), should decrease with a power of $K$.
We checked  
that the discrepancies vanish indeed like $1/K^2$, providing additional 
support for the approximate-vacuum conjecture.

\begin{table}[t]
\centerline{
\begin{tabular}{|c||c|c|c|c|}\hline
\rule[-3mm]{0mm}{8mm}
$2K$ & $M^2_{F_1}$ & $M^2_{F_2}$ & $M^2_{F_3}$ & $M^2_{F_4}$ \\
\hline\hline
3 &  4.5000 & --- & --- & --- \\ 
5 &  5.0000 & 12.5000 & --- & --- \\ 
7 &   5.2227 & 14.0000 & 24.5000 &  --- \\ 
9 & 5.3456  & 14.7645 & 27.0000 & 29.2451 \\ 
11 & 5.4222 & 15.2575 & 28.5484 & 32.0373 \\ 
13 & 5.4741 & 15.5908 & 29.6419 & 33.7443 \\ 
15 & 5.5111 & 15.8311 & 30.5931 & 34.7280 \\ 
17 & 5.5388 & 16.0113 & 30.3593 & 35.6396 \\ 
19 & 5.5602 & 16.1509 & 31.1301 & 36.3496 \\ 
21 & 5.5771 & 16.2618 & 31.6091 & 36.5054 \\ 
23 & 5.5908 & 16.3518 & 32.0304 & 37.0575\\ 
25 & 5.6021 & 16.4261 & 32.6060 & 37.4225 \\ 
27 & 5.6115 & 16.4884 & 32.0123 & 37.7243 \\ 
\hline
\rule[-3mm]{0mm}{8mm}
$\infty$ & 5.75 & 17.29 & 35.25 & 40.24 \\ 
\hline
\end{tabular}
}
\caption{Eigenvalues of the lowest four (suspectedly) single-particle states  
in the adjoint case. The masses are given in units $g^2N_c/\pi$.
The masses in the 't Hooft case at $K{=}b+\frac{1}{2}$ are 
{exactly} the same as in the bosonic sector at $K{=}b+1$. 
The mass of the only single-particle state in the large $N_f$ limit is
independent of the cutoff: $M^2_{N_f}(K)\equiv 1.0000 \frac{g^2N_f}{\pi}$.}
\label{Table2}
\end{table}

Furthermore, analyzing the 
the spectrum as a function of the continuous parameter $\lambda$,
it seems plausible that the number of single-particle states 
stays the same as in the 't Hooft case, see also the discussion next section.
This hypothesis can in principle be tested at $\lambda=1$ by using an
approximate multi-particle mass formula to eliminate the
multi-particle states.
The fact that most single-particle states 
asymptotically become multi-particle states at large $\lambda$ 
cannot affect us here.
At finite $\lambda$ the problem is to show that most states are 
multi-particle states, although they are not exactly following a
mass formula in the fashion of Eq.~(\ref{MultiFermions}).

Summing up the above findings, the conclusion is the following.  
Finding that the adjoint bosonic single-particle 
states do not form
multi-particle states and conjecturing
that the number of fermionic single-particle 
states grows linearly with $K$,
we get the same situation concerning
the ratio of single- to multi-particle states as in the 't Hooft case. 
We then conclude that 
the number of single-particle states is the same as in the 't Hooft model,
even in the bosonic sector.
Since the masses of the fermionic and bosonic single-particles 
are not degenerate as opposed to the 't Hooft case, 
we find two adjoint Regge trajectories, a bosonic and a fermionic one.
The determination of the functional dependence of the single-particle states
on the excitation number
requires further investigations
which are beyond the scope of the present work.

%
\begin{figure}
\centerline{
\psfig{file=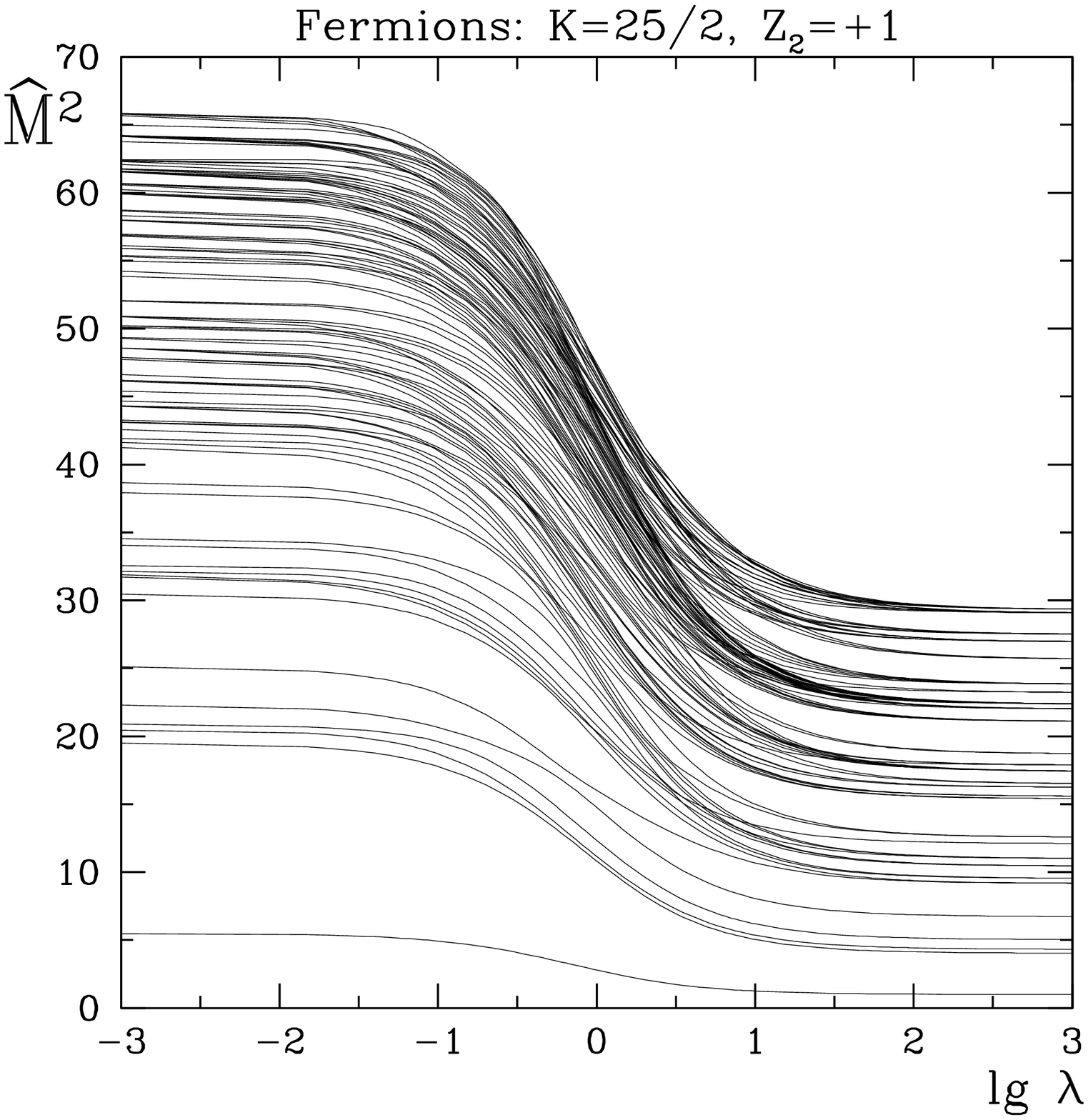,width=8cm}
\psfig{file=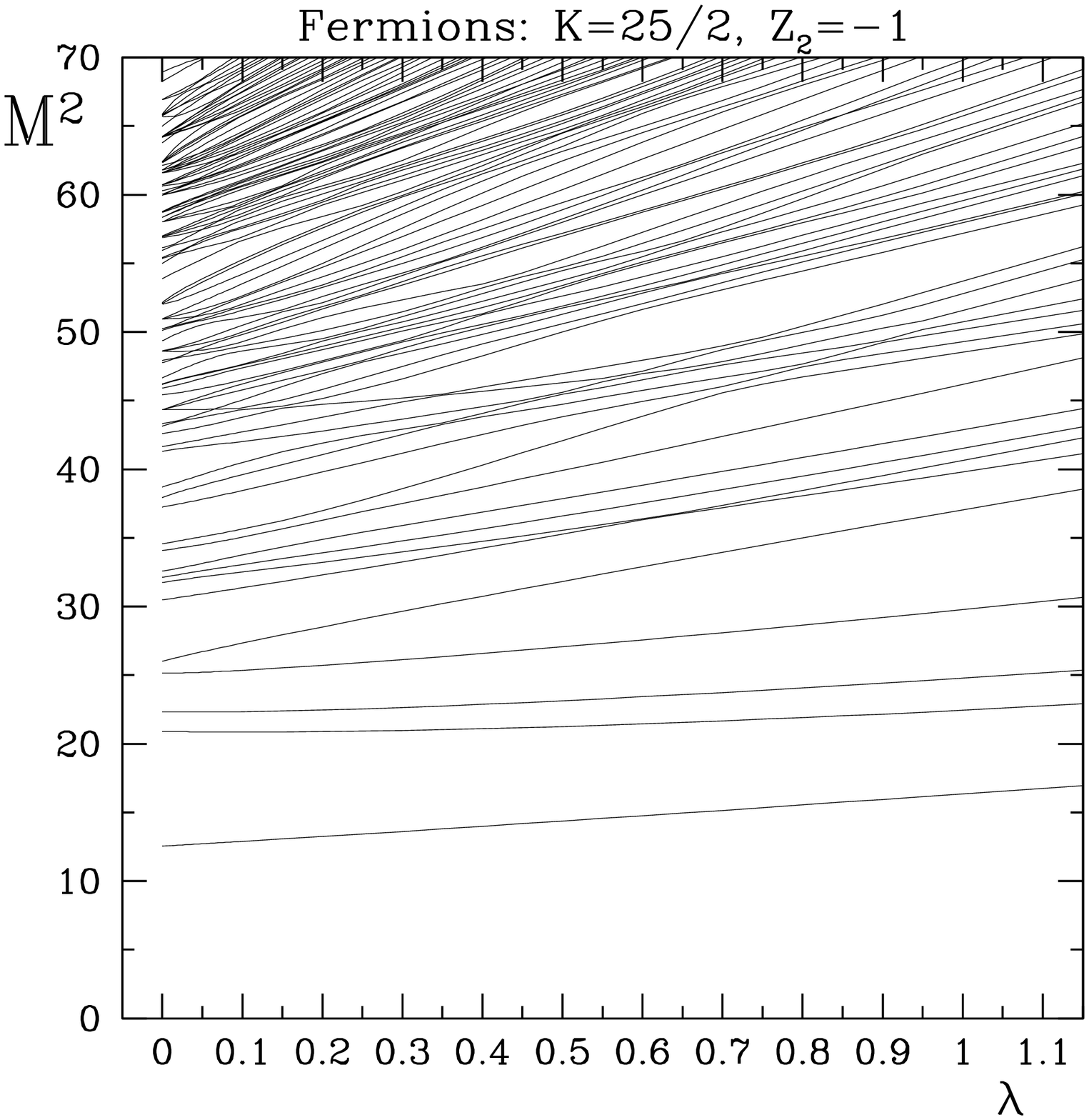,width=8cm}}
\centerline{
\psfig{file=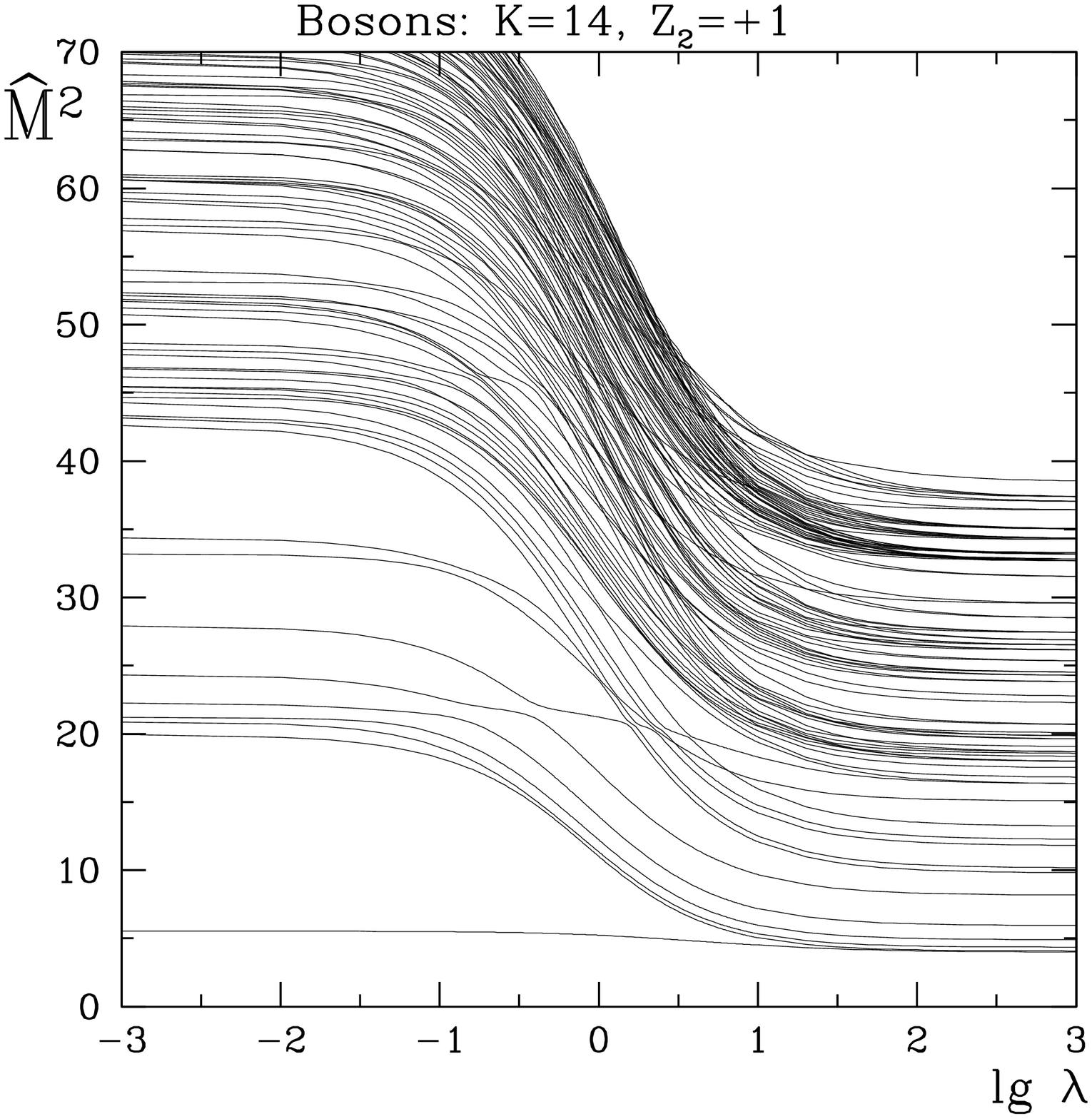,width=8cm}
\psfig{file=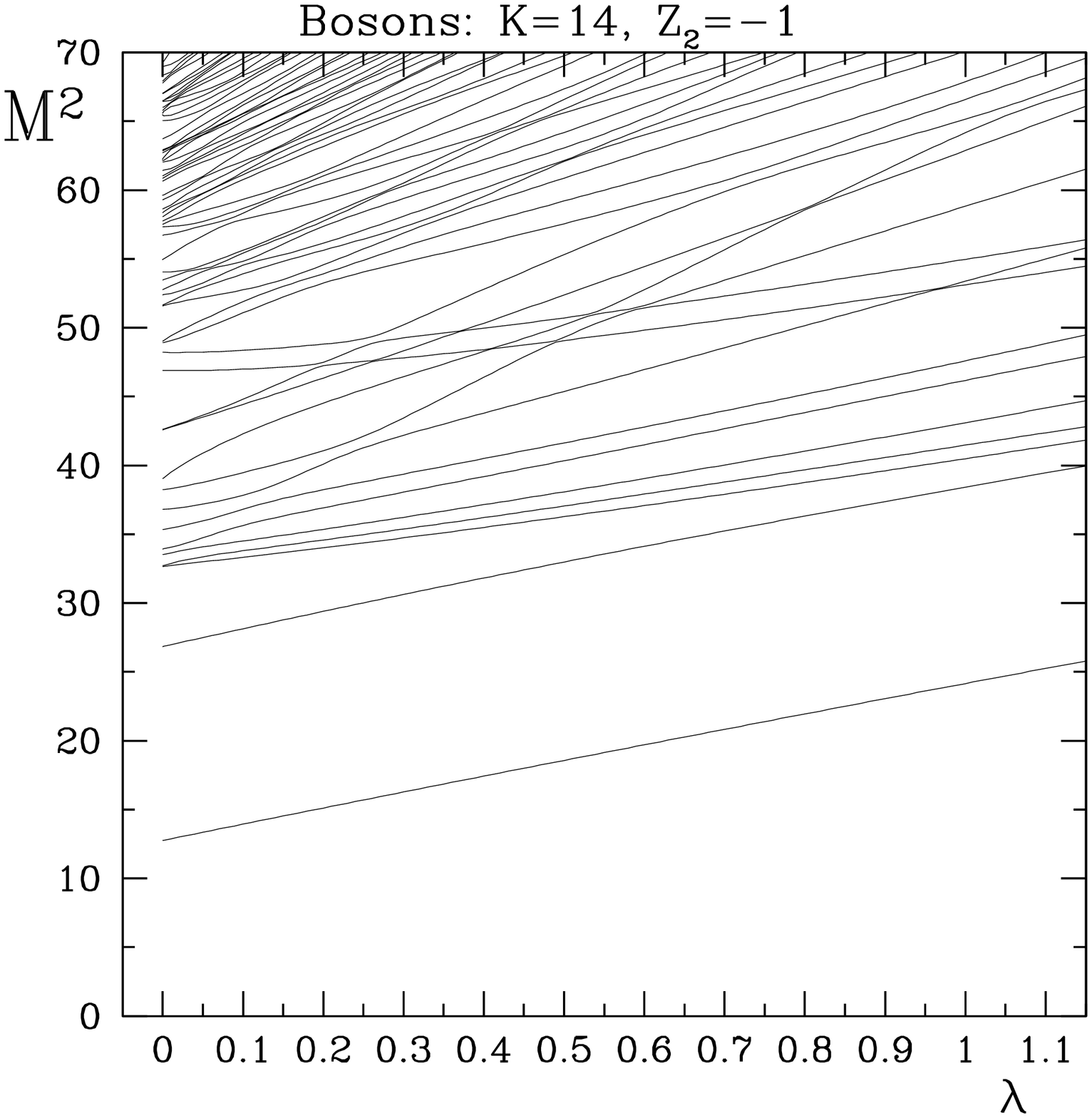,width=8cm}}
\caption{The spectrum of two-dimensional QCD in all sectors of the theory
as a function of $\lambda$. Plotted are the lowest 100 eigenvalues 
in the fermionic sectors (top row) and the bosonic sectors (bottom row).
Left column: $Z_2$ even sectors, 
reduced eigenvalues $\hat{M}^2\equiv M^2/(1+\lambda)$ vs.~$\lg\lambda$.
Right column: $Z_2$ odd sectors, actual eigenvalues vs.~$\lambda$.
\label{FullSpectrum}}
\end{figure}
%

\subsection{Intermediate cases and eigenfunctions}
\label{SecInter}

It is instructive to study the spectrum of eigenvalues as a function
of the continuous parameter $\lambda=N_f/N_c$. In   
Fig.~\ref{FullSpectrum} we plotted the spectrum of
two-dimensional Yang-Mills theories coupled to massless matter in 
representations characterized by $\lambda$,
in the Veneziano limit.
We show all sectors of the theory,
{\em i.e.}~the bosonic and fermionic $Z_2$ even and odd spectra, as a 
function of $\lambda$.
A crucial observation is that the 't Hooft mesons develop 
differently in the 
fermionic and bosonic sectors as $\lambda$ grows, and the
degeneracy of their masses is lifted. 
In particular, the mass of the lightest meson in the fermionic
sector decreases
as $\lambda$ grows with a slope of $\hat{e}^{(f)}_1(1)=-1.39$, see Appx.~A.
It reaches the minimum of its parabolic trajectory $M^2_{F_1}(\lambda)$  
at $\lambda=1/3$. Asymptotically it rises linearly with $\lambda$.
On the other hand, the mass of the lightest boson increases 
monotonously.
This scenario for small $\lambda$ 
is expected, since the mass of the lightest state in a theory
has to decrease in second order perturbation theory. 
It is evidence for the 
conjecture that the theory is incomplete if only 
its bosonic sector is considered \cite{Kutasov94},
because the lowest boson cannot be the lightest state of the full 
theory. 
In general, we obtain the 
first corrections in $\lambda$ to the 't Hooft meson masses 
in the bosonic sector
in complete agreement with the perturbative 
calculations by Engelhardt \cite{Engelhardt2001}, as we will show in 
more detail in Appx.~\ref{Appx:Engelhardt}. 

Concerning the global $\lambda$ dependence of the spectrum, 
Fig.~\ref{FullSpectrum},
we emphasize that the eigenvalues are smooth functions of $\lambda$, and
we seem to find no indication that the adjoint theory is special.
We see a lot of level crossings, some of which are obscured 
by eigenvalue repulsion due to finite harmonic resolution.
Some of the reduced eigenvalues $\hat{M}^2\equiv M^2/(1+\lambda)$ 
are almost stationary as a function of the parameter
$\lambda$, amongst them chiefly the suspected single-particle states.
Note, however, the somewhat artificial definition of the reduced masses,
which was used in order to fit the spectrum for all values of $\lambda$ 
into one plot.

\begin{figure}
\centerline{
\psfig{file=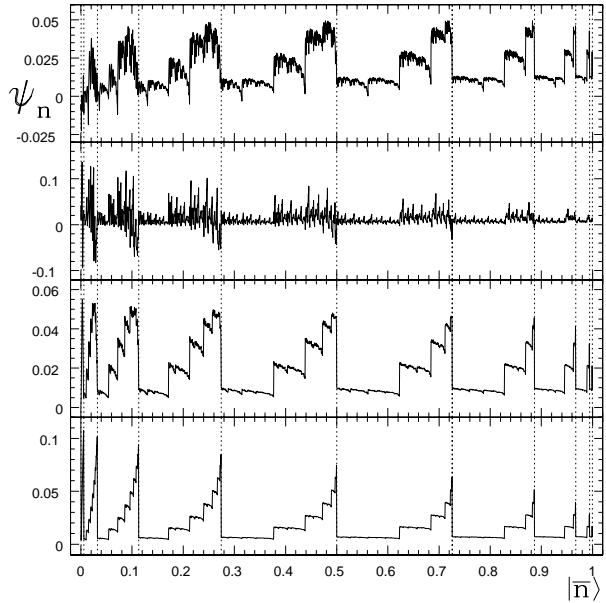,width=8cm}}
\caption{The wavefunctions of four adjoint states at $K=25/2$:
(a) $M^2=5.6021$, (b) $M^2=16.4261$, (c) $M^2=21.8688$, (d) $M^2=32.6060$
[from bottom to top].
Plotted are the amplitudes $\psi_n$ vs.~$\bar{n}=n/2^{K-3/2}$.
The third state is the only multi-particle state in this plot 
and its eigenfunction has clearly a different shape.
The number of currents in a basis state changes at the dashed lines.
\label{wfs}}
\end{figure}

It would be very interesting if one could find a criterion 
for a state to be a single-particle state, or if one could 
formulate a good observable, {\em e.g.}~a structure function, that
would allow to distinguish single- from multi-particle states.
We display four adjoint eigenfunctions in Fig.~\ref{wfs}. 
Apart from the striking repetitive pattern in the different 
parton sectors, we see that the multi-particle state in this 
plot is distinct from the single-particle states.  
In the extreme cases, $\lambda=0$ and $\lambda\rightarrow\infty$, 
we obtain the following behavior of the wavefunctions.
The 't Hooft eigenfunctions are very similar to the ones in the adjoint case.
They can in principle be calculated from the 
standard formulation of the theory with fermion fields, which is 
equivalent to a change of basis. 
At large $N_f$ the wavefunctions are converging very slowly. They 
look very much like in the 't Hooft limit for 
$K{<}10$. At that point, most of the amplitudes become suppressed 
while keeping their shape, and the 
amplitudes of states with a large number of currents become 
heavily peaked.

%
%

\vspace{-0.1cm}
\section{Summary and Discussion}
\label{SecSummary}

In this article we 
presented the spectrum of two dimensional Yang-Mills 
theories coupled to massless matter in a representation 
characterized by the ratio of the numbers of 
flavor and color $\lambda$ 
in the Veneziano limit.
We derived the Hamiltonian in the fermionic sector in the framework 
of DLCQ as an algebraic 
function of the harmonic resolution $K$ and the ratio $\lambda$. 
Surprisingly, we found the momentum of an adjoint fermion depending
on $\lambda$ in this discrete approach. 
This is explained by the fact that we have zero modes of the
current operators in the theory, while fermionic zero modes are recovered 
in the continuum limit only.
The well-known spectra in the 't Hooft and the large $N_f$ limits were
reproduced. 
Although this is not surprising taking into account the universality
established in Ref.~\cite{KutasovSchwimmer95} which is a   
specialty of two dimensions, it is nevertheless a strong check on the
numerics. 
We found the bosonic and fermionic spectra
to be degenerate in the 't Hooft limit, and the only meson 
of the large $N_f$ limit in the fermionic sector. 
The multi-particle states decouple completely in these limits and
a construction of the spectra in terms of their single-particle content
was achieved. This allowed for a complete classification of all states 
including their statistics and symmetry properties in both cases.
In trying to apply this knowledge to the adjoint case,
we were only partly successful.
We presented evidence for the {\em conjecture} that the vacuum is 
only approximately realized in the DLCQ formulation of the fermionic sector
of the theory.
This conjecture allowed us to understand the empirical finding that 
the multi-particle states have only fermionic single-particle constituents.
This fact was deduced from an analysis of the spectra at 
intermediate values of $\lambda$. 
The approximate vacuum induces couplings between single- and multi-particle 
states which were found to decrease with the harmonic resolution like 
$1/K^2$, {\em i.e.}~consistent with the above conjecture.
We motivated the {\em hypothesis} that the number of fermionic 
single-particle states is the same in the 't Hooft and adjoint cases
by pointing out the smooth transition of the spectra into each other
by the continuous parameter $\lambda$.
We then concluded that there has to be a second Regge trajectory of 
bosonic single-particle states, because at each $K$ 
the size of the Fock basis
and the number of multi-particle states determined by 
the kinematics of their fermionic constituents allows for exactly
$K-1$ additional states.
Although we were as of yet unable to give the complete solution of the 
adjoint theory in terms of its single-particle states, it seems thus that 
their number grows linearly with the harmonic resolution. Their 
masses tend to grow more rapid with the excitation number $n$,
maybe like $M^2\propto n^2$, rather than linear as in the 't Hooft model. 

The two Regge trajectory conjecture is not in contradiction
with the expectation of a multi-Regge structure
at non-vanishing fermion mass $m$ \cite{Kutasov94},
or with the related appearance of a 
Hagedorn spectrum signaling the confinement/screening
transition as $m$ vanishes \cite{BoorsteinKutasov97}.  
When a fermion mass is turned on, the description of the 
theory in terms of Kac-Moody currents breaks down and
the theory turns from screening into a confining phase. 
It has been pointed out by Kutasov how in the massless theory the 
two seemingly contradictory facts of having a vanishing 
string tension together with the absence of a Hagedorn transition,
can be reconciled \cite{BoorsteinKutasov97}. 
In short, the exponentially rising density of states characteristic of a
Hagedorn transition can be explained
by a large degeneracy in the massless sector of the theory
which is lifted if a fermion mass is turned on.
This sector is, of course, completely inaccessible in the 
present approach.

%

In summary, we hope to have added some new pieces of information 
to the adjoint QCD$_2$ puzzle. 
The major practical goal remains to identify all single-particle 
states of the theory unambiguously. 
While we were unable to present a complete solution,
we described a practical algorithm
to extract the single-particle spectrum. Using  
the spectral information presented here, one could 
identify all multi-particle states 
by the characteristic $K$ dependence of their masses 
dictated by an approximate multi-particle mass formula. 
While this seems in principle possible, it is nevertheless 
beyond the scope of the present work.
This exercise can also serve as a 
quantitative test of the hypothesis that the number of 
single-particles is the same in the 't Hooft and adjoint models.
Furthermore, the approximate-vacuum conjecture can be ruled out, if one could
show that the deviations from the discrete multi-particle mass formula 
do not fall off everywhere with the resolution $K$.
Improvements of the results might be possible 
by attacking the theory from a very different point.
The adjoint theory is supersymmetric at a specific value 
of the fermion mass $m=g\sqrt{N_c}$
\cite{BoorsteinKutasov97}.
In the light of recent progress in the evaluation of supersymmetric 
theories \cite{SDLCQ}, this might be an interesting alternative.
%


\begin{appendix}

\section{Recovering the corrections to the 't Hooft masses}
\label{Appx:Engelhardt}

\begin{figure}
\centerline{
\psfig{file=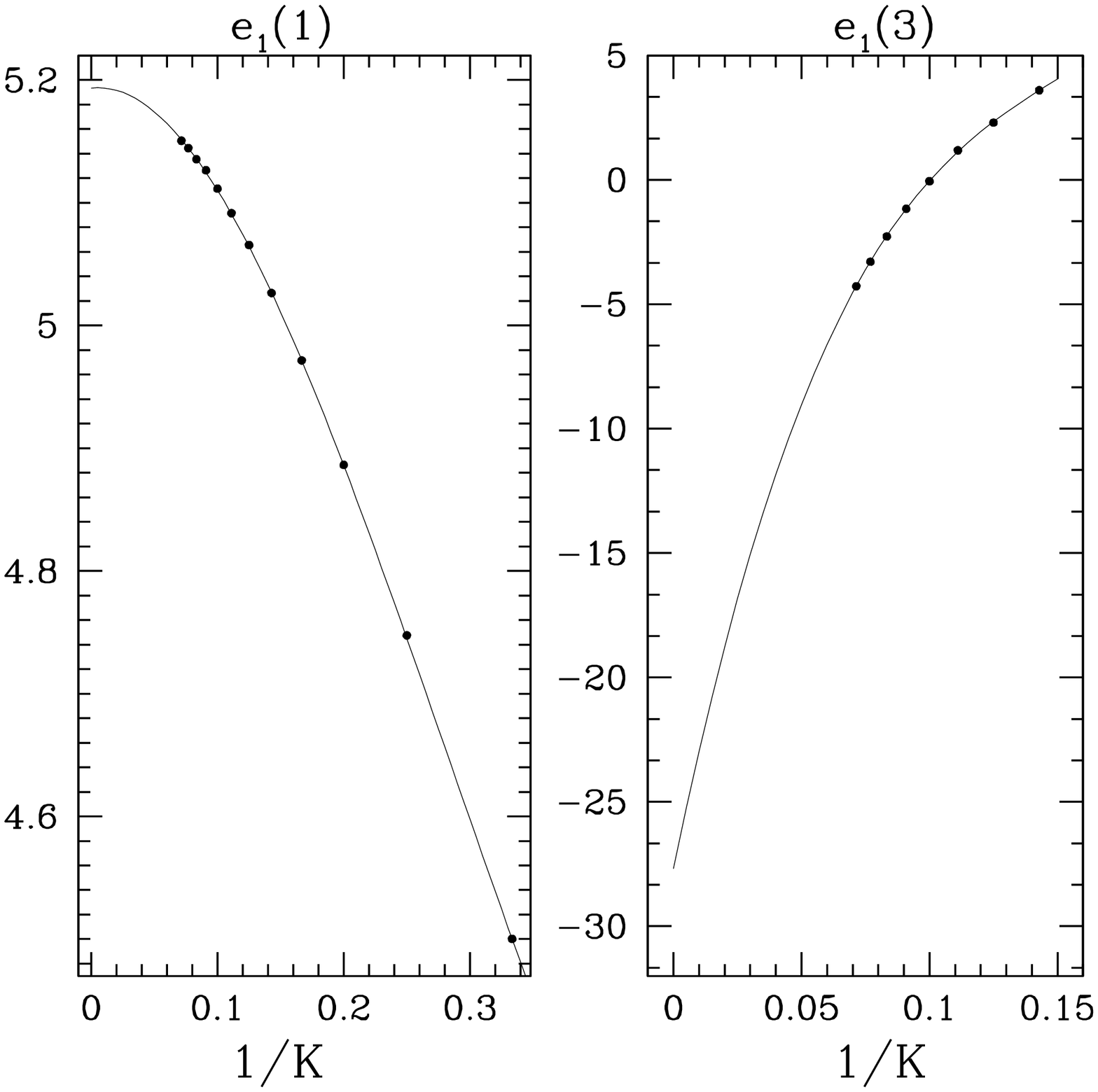,width=7cm}
\psfig{file=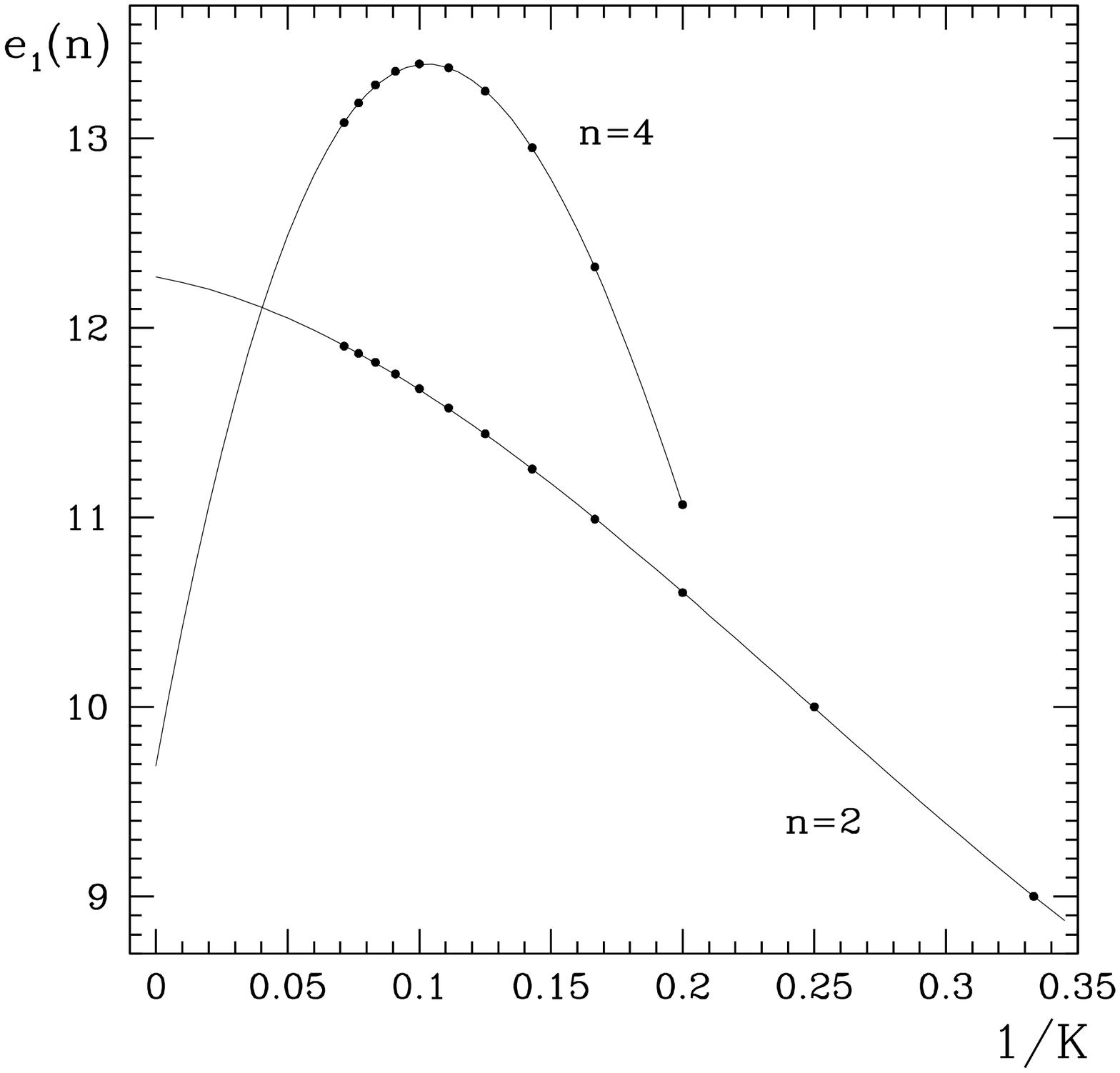,width=7cm}}
\caption{
The coefficients $e_1(n)$ of the first correction in $\lambda$
to the 't Hooft meson masses $M^2_n(\lambda=0)$ vs.~$1/K$. 
\label{fig1}}
\end{figure}

In a recent paper, Engelhardt \cite{Engelhardt2001} calculated the 
first corrections in $\lambda=N_f/N_c$ to the masses of the lowest four 't Hooft
mesons, namely the slopes $e_1(n)$, $n=1,2,3,4$ in the 
expansion 
\[
M^2_n(\lambda)=M^2_n(0)+e_1(n) \lambda +\ldots, \label{expansion}
\] 
where the masses are in units $g^2N_c/\pi$.
Surprisingly, some of these corrections are
negative and large. This seems to contradict the results of Ref.~\cite{UT},
Fig.~5(b), and we shall re-analyze them here. 
The slope of the curves $M^2_n(\lambda)$ is 
strongly dependent on the harmonic 
resolution $K$ for small $\lambda$. 
If we plot the slopes as a function of $1/K$ we see a consistent 
picture arising.
We fitted the slopes for the lowest four 't Hooft mesons, 
Fig.~\ref{fig1}, to a polynomial of third order in $1/K$, and 
obtain in the continuum limit   
\begin{equation}
\hat{e}_1(n)= 5.19, 12.27, -27.7, 9.69,
\end{equation}
to be compared to Engelhardt's values 
\begin{equation}
e_1(n)= 5.1, 12.0, -30.5, 9.1.
\end{equation}
Note that Engelhardt's values are lower
bounds (although one expects very small corrections), whereas 
extrapolations towards the continuum 
in DLCQ tend to be upper bounds. In this sense, the agreement is fairly well.

We emphasize that for small $K$ one is totally misled as to what 
the continuum limit might be for the slopes $e_1(3)$ and $e_1(4)$, 
{\em cf.}~Fig.~\ref{fig1}.
For instance the slope of the third state,
$e_1(3)|_{K\rightarrow\infty}=-27.7$, is
still positive at the fairly large resolution $K=10$.
This shows the importance of Fock states with a large number of currents, and
renders two-current approximations questionable. 
Note that the mass of this state increases linearly for large enough $\lambda$,
although it starts with a large negative slope,
{\em cf.}~Fig.~5(b) of Ref.~\cite{UT}. 
The first correction to the 't Hooft mass is a good approximation 
only up to $\lambda\simeq 0.01$. 

It is important to note that the corresponding (and degenerate) 't Hooft 
single-particle states in the 
fermionic sector have different corrections in $\lambda$.
Following the development of the three 
lowest 't Hooft mesons in the 
fermionic sector we obtain the following slopes
\begin{equation}
\hat{e}^{(f)}_1(n)= -1.36, 1.94, -15.38.
\end{equation}
The lowest 
state, which develops into the lightest adjoint 
state as $\lambda$ grows to unity, has a negative correction, as expected from
second order perturbation theory for the lowest state of a theory.
It is thus clear that the single-particle states in the fermionic and 
bosonic sectors are distinct entities, although their masses are
degenerate in the 't Hooft limit.
The fourth lightest state 
has a rather irregular trajectory $\hat{e}_1(4,K)$
which prevents us from extrapolating to the continuum.
This is easily understood when one compares the spectra in the bosonic
and the fermionic 't Hooft sectors. In the bosonic sector, all four lowest 
single-particle states are lighter than the lowest multi-particle states
in their sector. In the fermionic sector the fourth single-particle state 
lies in the  two-particle continuum formed by the lightest massive
't Hooft meson
\footnote{The lightest 't Hooft meson proper
with $M^2(n=0)=0$ is absent from the spectrum, 
since we work formally in the adjoint theory. 
}.
In perturbation theory, such a situation has to be taken care of by 
constructing states which contain admixtures of degenerate 
states with higher parton numbers \cite{Engelhardt2001}. 
We see that in the present discrete approach we run into 
the same difficulties.

%
%

\end{appendix}


\vspace{0.4cm}
\centerline{\large\bf Acknowledgments}
\vspace{0.1cm}

The author thanks S.S.~Pinsky and J.R.~Hiller 
for careful reading of the manuscript.
Discussions with M.~Engelhardt about his results, Ref.~\cite{Engelhardt2001},
are acknowledged. 
This work was supported by an Ohio State University Postdoctoral Fellowship.





\end{document}